\documentclass[a4paper,10pt]{article}

\usepackage[utf8]{inputenc}

\usepackage{authblk}

\usepackage{ifthen,ifpdf}

\ifpdf
\usepackage{hyperref}
\hypersetup{colorlinks=false,allcolors=blue}
\usepackage{hypcap}
\pdfpagewidth=\paperwidth
\pdfpageheight=\paperheight
\fi

\usepackage[utf8]{inputenc}
\usepackage[T1]{fontenc}

\usepackage{amsmath, amsthm, amssymb}
\usepackage{mathtools}
\usepackage{enumerate}
\usepackage{afterpage}
\usepackage{tabularx}
\usepackage{dsfont}

\usepackage{stmaryrd}

\usepackage{graphicx}

\usepackage{algorithm}
\usepackage{algpseudocode}
\algtext*{EndWhile}%
\algtext*{EndIf}%
\algtext*{EndFor}
\algtext*{EndProcedure}

\usepackage{caption}
\usepackage[labelformat=simple]{subcaption}

\usepackage{color}
\usepackage{array, multirow}
\usepackage{booktabs}

\newcolumntype{M}[1]{>{\vspace{3pt}\raggedleft\arraybackslash}m{#1}}

\usepackage{siunitx}[]
\usepackage{tikz}
\usetikzlibrary{calc}%
\usetikzlibrary{matrix}
\usepackage[export]{adjustbox}
\usepackage{pgfplots}
\usepackage{pgfplotstable}
\usepgfplotslibrary{units}
\pgfplotsset{compat=1.9,samples=1000,every axis/.append style={
		font=\large,
		line width=1pt,
		tick style={line width=0.8pt}}}

\usetikzlibrary{arrows}
\usetikzlibrary{hobby}
\usetikzlibrary{calc}
\usetikzlibrary{arrows}
\usetikzlibrary{shapes}
\usetikzlibrary{positioning}
\usetikzlibrary{intersections}
\usetikzlibrary{decorations.pathreplacing}
\usetikzlibrary{patterns}
\usetikzlibrary{decorations.pathmorphing}
\usetikzlibrary{decorations.markings}
\usetikzlibrary{3d}
\usetikzlibrary{matrix}
\usetikzlibrary{spy}

\usepgfplotslibrary{polar}
\usepgfplotslibrary{external}
\usepgfplotslibrary{fillbetween}
\usepgfplotslibrary{groupplots}
\usepgfplotslibrary{patchplots}

\usepackage{cite}

\usepackage{anysize} 
\marginsize{2.5cm}{2.5cm}{2cm}{2cm}

\usepackage{wrapfig}

\usepackage{upgreek}
\theoremstyle{plain}

\theoremstyle{remark}

\numberwithin{equation}{section}

\usepackage{bm}
\usepackage[scr=dutchcal]{mathalpha}
\usepackage{overpic}

\newcommand{\R}{\mathds{R}}

\newcommand*\diff{\mathop{}\!\mathrm{d}}

\newcommand{\Sym}[1]{\textrm{Sym}(#1)}

\renewcommand{\div}{\textrm{div }}

\makeatletter
\newcommand{\proplabel}[1]{\protected@edef\@currentlabel{(#1)}\phantomsection\label{prop:#1}}
\makeatother
\newcommand{\propref}[1]{\textbf{\ref{prop:#1}}}
\newcommand{\parhead}[1]{\par\medskip\noindent\underline{#1}:\ }

\DeclareMathOperator{\eps}{\varepsilon}

\newcommand{\styy}[1]{{\mathbb{#1}}}

\newcommand{\fd}{\bm{ d}}

\newcommand{\fn}{\bm{ n}}

\newcommand{\fu}{\bm{ u}}
\newcommand{\fv}{\bm{ v}}
\newcommand{\fw}{\bm{ w}}
\newcommand{\fx}{\bm{ x}}

\newcommand{\bfeps}{\bar{\feps}}

\newcommand{\feps}{\bm{\varepsilon}}
\newcommand{\fsigma}{\mbox{\boldmath $\sigma$}}

\newcommand{\fsymgrad}{\mbox{\boldmath $\nabla^s$}}

\newcommand{\ffC}{\styy{ C}}

\newcommand{\T}[1]{{#1}^{\sf  T}}

\newcommand{\norm}[1]{\|#1\|}

\newcommand{\fzero}{\bm{ 0}}

\newcommand{\invfourier}[1]{\mathcal{F}^{-1}\left(#1\right)}
\newcommand{\fourier}[1]{\mathcal{F}\left(#1\right)}

\newcommand{\displacement}{d}
\newcommand{\sector}{\mathscr{s}}
\newcommand{\averageStrainMatrix}{\overline{\mathbf{\Upsilon}}}

\title{Scaled boundary finite elements\\ for FFT-based computational micromechanics}

\author[1]{Flavia Gehrig}
\author[2]{Carolin Birk}
\author[1,3,4,*]{Matti Schneider}

\affil[1]{University of Duisburg-Essen, Institute of Engineering Mathematics, Essen}
\affil[2]{University of Duisburg-Essen, Institute of Structural Analysis, Essen}
\affil[3]{Center for Nanointegration Duisburg-Essen (CENIDE), Duisburg}
\affil[4]{Fraunhofer Institute for Industrial Mathematics ITWM, Kaiserslautern}
\affil[*]{correspondence to: \texttt{matti.schneider@uni-due.de}}

\date{\today}

\begin{document}
\maketitle 
\begin{abstract}
\noindent
We introduce a matrix-free FFT-based solver for three-dimensional linear elastic homogenization problems discretized by the scaled boundary finite element method (SBFEM) on a voxel grid. Owing to its analytical solution in the radial direction, the SBFEM discretization is inherently softer than fully integrated voxel finite elements, whereas its displacement-based formulation remains symmetric, conforming, and supports arbitrary stiffness anisotropy. These properties make SBFEM an attractive candidate for porous and highly heterogeneous microstructures, where existing voxel-based FFT discretizations offer no generally reliable choice: Reduced-integration hexahedral elements, equivalently the rotated staggered grid discretization, may suffer from hourglass instabilities in the presence of voids, whereas fully integrated hexahedral elements are often too stiff.\\
The proposed discretization uses hexahedral SBFEM cells on the full voxel mesh of a regular periodic grid, retaining the simplicity of a fixed-resolution voxel representation without requiring adaptive mesh coarsening. The main contribution is a preconditioner based on the homogeneous SBFEM stiffness operator. We show that this preconditioner may be applied locally in Fourier space and that the condition number of the preconditioned system is bounded by the material contrast, independently of the grid resolution. This result enables a fully matrix-free implementation of the SBFEM-discretized problem, without assembling or storing the global stiffness matrix.\\
Numerical experiments on a short-fibre reinforced composite, a polyurethane foam and a sand-based binder jet microstructure demonstrate robust convergence and competitive accuracy compared with established voxel discretizations. In particular, SBFEM remains stable for porous microstructures, for which reduced-integration hexahedral elements become unstable.\\
	\quad\\
	{\noindent\textbf{Keywords:} FFT-based homogenization; scaled boundary finite element method; SBFEM-FFT preconditioner; voxel-based discretization; porous materials; matrix-free solvers}
\end{abstract}

\newpage

\newpage

\section{Introduction}
\label{sec:intro}

Components made from heterogeneous materials, such as fibre-reinforced composites, porous materials or polycrystals, feature microstructures whose characteristic length is much smaller than the dimensions of the component. Resolving the microstructure in a component-scale simulation is prohibitively expensive in general. Homogenization theory~\cite{bensoussan1978asymptotic,sanchez1980non,milton2002theory} replaces the heterogeneous material by an equivalent homogeneous material, whose effective properties follow from a cell problem posed on a representative volume element (RVE) of the microstructure. For linear elastic materials, the effective behaviour is characterized exactly by an effective stiffness tensor, which is determined by solving the cell problem for six linearly independent load cases~\cite{michel1999effective,milton2002theory}. For nonlinear materials, in contrast, no such finite-dimensional characterization is available in general, and concurrent multiscale schemes solve a cell problem at every macroscopic integration point~\cite{feyel2000fe2,geers2010multi,matous2017review}. In both cases, the cell problems are solved by full-field simulations on the RVE, which calls for efficient and robust solvers.

Fast Fourier Transform (FFT)-based computational homogenization methods, pioneered in the 1990s by Moulinec and Suquet~\cite{moulinec1994fast,moulinec1998numerical}, provide a highly efficient and robust framework for such simulations. These methods avoid the need for explicit matrix assembly, significantly reducing memory consumption and computational cost. By leveraging state-of-the-art FFT algorithms~\cite{frigo2005design,dalcin2019fast} in an iterative solution scheme, these methods require a computational effort of order \mbox{$\mathcal{O}(N\log N)$} per iteration, where $N$ denotes the number of voxels, and their iteration count does not grow with the resolution. Thus, they remain efficient for complex, high-resolution microstructures. Their use of regular grids circumvents expensive mesh generation and enables direct application to in-situ microstructures measured by modern imaging techniques~\cite{chen2019analysis,rovinelli2020assessing}. Periodic boundary conditions, which are particularly suitable to minimize the bias in the effective properties of an RVE~\cite{terada2000simulation,van2000overall,kanit2003determination}, are natural in FFT-based homogenization and do not require any additional effort. Nevertheless, extensions towards Dirichlet and Neumann boundary conditions are possible~\cite{gelebart2020modified,sancho2023implicit,risthaus2024imposing}.

A key insight for the development of modern FFT-based methods was the separation of the original method~\cite{moulinec1994fast,moulinec1998numerical} into \emph{discretization} and \emph{solution scheme}~\cite{brisard2012combining,willot2014fourier,vondvrejc2014fft,schneider2015convergence}: The original method is based on an equivalent reformulation of the (quasi-)static equilibrium condition as a Lippmann-Schwinger equation~\cite{kroner1972statistical,zeller1973elastic}, which is then solved by an iterative scheme involving the FFT. In the 2010s, the underlying discretization of the original method was identified as a trigonometric collocation discretization method~\cite{nguyen2011efficient} and a nonconforming Fourier-Galerkin method~\cite{vondvrejc2014fft}. The underlying iterative solution scheme, known as the basic scheme, was interpreted as a gradient descent method~\cite{kabel2014efficient} or Richardson iteration~\cite{mishra2016comparative}. With these insights, it was possible to further increase the speed of FFT-based methods by importing solution schemes from large-scale optimization~\cite{zeman2010accelerating,schneider2019barzilai,wicht2020quasi}. Moreover, the polarization-based schemes~\cite{eyre1999fast,michel2001computational}, which were originally conceived specifically for FFT-based methods, were identified as Douglas-Rachford splitting methods and as the alternating direction method of multipliers (ADMM)~\cite{schneider2019polarization}. For an overview of the characteristics of these solution schemes, we refer to chapter 3 of the review article~\cite{schneider2021review}. Moreover, alternative discretizations based on the Hashin-Shtrikman principle~\cite{brisard2010fft,brisard2012combining} or the Fourier–Galerkin approximation~\cite{bonnet2007effective,vondvrejc2015guaranteed} were developed and existing voxel-based discretizations using finite differences~\cite{willot2015fourier,schneider2016computational,finel2025tetrahedron}, finite volumes~\cite{dorn2019lippmann} and finite element methods (FEM)~\cite{schneider2017fft,leuschner2018fourier,ladecky2023optimal} were adapted to the FFT-based framework.

Adapting a discretization to the FFT-based framework includes deriving a suitable preconditioner, which
\begin{enumerate}
    \item[\textbf{(a)}]\proplabel{a} may be applied locally in Fourier space, i.e., independently for each frequency, as is the case for operators with homogeneous coefficients on a regular grid, \emph{and}
    \item[\textbf{(b)}]\proplabel{b} leads to a well-conditioned system with stable iteration count, for example, by providing a mesh-independent upper bound on the condition number of the preconditioned system.
\end{enumerate}
Recently, higher accuracy discretizations based on composite voxels~\cite{gelebart2015filtering,kabel2015use,lendvai2024assumed,keshav2023fft}, the strongly stable extended/generalized finite element method~\cite{gehrig2025x,gehrig2026stable}, isogeometric analysis~\cite{montardini2025isogeometric} and mesh morphing strategies~\cite{zecevic2022new,bellis2024numerical,zecevic2026achieving,bignonnet2026fourier} were introduced to FFT-based methods. However, so far, only the composite voxels~\cite{gelebart2015filtering,kabel2015use,lendvai2024assumed,keshav2023fft}, the strongly stable extended/generalized finite element method~\cite{gehrig2025x,gehrig2026stable} and the isogeometric analysis~\cite{montardini2025isogeometric} were demonstrated to reliably fulfil properties 
\propref{a} and \propref{b} for the preconditioner. Mesh morphing strategies still struggle to satisfy property \propref{b} in their preconditioning strategy, which results in an iteration count that increases with increasing resolution~\cite[Fig.~7b]{bignonnet2026fourier}.

Provided that a preconditioner with the properties \propref{a} and \propref{b} may be developed, additional voxel discretizations may be incorporated into the FFT-based framework. Such an additional voxel discretization is required for porous and highly heterogeneous microstructures, for which none of the established voxel discretizations is generally reliable. Discretizations based on trigonometric polynomials, both the Fourier-Galerkin discretization~\cite{vondvrejc2014fft} and its underintegrated variant, the original Moulinec-Suquet discretization~\cite{moulinec1994fast,moulinec1998numerical}, do not lead to convergent solvers for infinite material contrast~\cite{schneider2016computational,schneider2020lippmann}. The same holds for finite-difference discretizations with wide stencils~\cite{schneider2021review,dave2026limitations}. The staggered grid discretization~\cite{schneider2016computational} leads to a non-symmetric system, and the tetrahedron-based discretization~\cite{finel2025tetrahedron} is prone to locking. Reduced-integration hexahedral elements, equivalently the rotated staggered grid~\cite{willot2015fourier,schneider2017fft}, may suffer from hourglass instabilities in the presence of voids~\cite{schneider2020lippmann}, whereas fully integrated hexahedral elements lead to an overly stiff response at coarse resolutions~\cite{schneider2022voxel,gehrig2025element}.
The scaled boundary finite element method is a candidate to close this gap. Owing to its semi-analytical approximation, the SBFEM is softer than fully integrated hexahedral elements, and it retains a symmetric, conforming, displacement-based formulation that supports arbitrary anisotropic stiffness tensors.

The SBFEM was originally proposed as a semi-analytical technique in the context of soil-structure interaction~\cite{SongW-97}. It is based on a coordinate transformation that facilitates the separation of local coordinates describing the boundary of the domain and a radial coordinate defined with respect to a point from which the complete boundary must be visible. Using this transformation together with a semi-discretization of the unknown field in, for example, a virtual work statement, yields an ordinary differential equation for the variation of the unknown field in the radial direction. For elastostatics, this ordinary differential equation may be solved in closed-form. For dynamic problems in unbounded domains and fracture problems, this semi-analytical approach yields rigorous representations of radiation damping and stress singularities, respectively. Moreover, the SBFEM facilitates the construction of polytope elements since it requires only the discretization of the boundary. Polygon elements provide great flexibility in meshing and re-meshing. For the above reasons, the SBFEM in its original form was used in various contexts, including acoustics~\cite{Lehmann2006,Liu2019,Khajah2021}, soil-structure interaction~\cite{Bazyar2006,Chen2014,Chen2015,Kuhn2024}, electromagnetic waveguides~\cite{Lin2011,Liu2012}, piezoelectricity~\cite{Mayland2009,Li2013}, wave propagation~\cite{Gravenkamp2015,Gravenkamp2017,Zhang2021}, discrete crack propagation modelling~\cite{Ooi2012,Ooi2015}, and adaptive phase-field modelling~\cite{Hirshikesh2019,Assaf2022}, for example.

On the other hand, the semi-analytical solution approach also limits the applicability of SBFEM in its original form to certain types of problems. In the last decade, the SBFEM was generalized by using it to construct shape functions on polytopes that may be used analogously to FE shape functions in the context of nonlinear analysis and for coupled problems~\cite{Chiong2014,Xiao2023}. This development facilitated applications of SBFEM to, for example, functional grading~\cite{Chiong2014}, elastoplasticity~\cite{Ooi2014}, poroelasticity~\cite{Ooi2018,Pramod2024}, thermoelasticity~\cite{Ooi2020,Iqbal2023} and finite strain~\cite{Ooi2025}, where further generalization was achieved by developing an isoparametric concept for polygon elements. For a recent review of papers addressing integration of SBFEM with automated mesh generation techniques and high-performance computing, we refer to Zhang and Song~\cite{ZhangS2026}. 

Applications of SBFEM to homogenization problems may be found in the literature~\cite{wang2017efficient,talebi2019scaled,saputra2020three,Tian2024}. Here, polygon meshes or hierarchical meshes (quadtree and octree meshes) are used to efficiently approximate the respective microstructural geometries and direct solvers are typically employed.

In these works, the hierarchical and polygonal meshes primarily serve to reduce the number of degrees of freedom to such an extent that direct solvers remain feasible. In particular, Gaddam et al.~\cite{gaddam2025octree} compared an octree-based SBFEM discretization, solved by a sparse direct solver, with the FFT-based spectral solver of DAMASK~\cite{roters2019damask}. FFT-based methods, in contrast, rely on iterative solvers and treat regular grids with billions of degrees of freedom, such that a reduction of the degrees of freedom is not necessary. On a regular grid, the SBFEM resolves the voxel image at its native resolution and is thus at least as accurate as on a coarsened mesh derived from the same image. Whether the SBFEM discretization itself may be solved by FFT-based methods, i.e., whether a preconditioner with the properties \propref{a} and \propref{b} exists for the SBFEM, is an open question.
This work introduces a voxel-based SBFEM discretization for FFT-based linear elastic homogenization and answers this question affirmatively.\\
The first contribution is a full-voxel SBFEM discretization for periodic three-dimensional linear elastic homogenization problems. In contrast to previous SBFEM approaches to homogenization, which rely on octree coarsening or polygonal aggregation, we deliberately retain the full voxel mesh. This choice keeps the discretization simple, avoids the geometric and implementation overhead associated with adaptive coarsening, and provides the full geometric fidelity of the voxel image.\\
The second and central contribution is an SBFEM-FFT preconditioner. FFT-based solvers rely on a homogeneous reference problem, which needs to be uniformly equivalent to the actual problem, i.e., with equivalence constants independent of the grid resolution. For classical discretizations, the discrete space does not depend on the material, and the equivalence follows directly from the bounds on the material stiffness. For the SBFEM, in contrast, the discrete space changes with the material, since the radial approximation is adapted to the local stiffness. Thus, the reference problem and the actual problem are posed on different discrete spaces, and their equivalence is not clear a priori. We show that the equivalence holds nevertheless, since both condensed stiffness operators result from minimization over the same material-independent semi-analytical space. Consequently, the homogeneous SBFEM stiffness operator serves as the preconditioner, which admits a local representation in Fourier space (property \propref{a}) and may be applied by matrix-valued Fourier multipliers, without assembling a global matrix.\\
The third contribution is the resulting spectral estimate for the preconditioned system. We provide the arguments that the condition number is bounded from above by the material contrast and is therefore independent of the grid resolution (property \propref{b}). Thus, the SBFEM discretization may be combined with modern FFT-based solvers with stable iteration counts, just as the classical voxel discretizations.\\
The fourth contribution is a numerical assessment of the resulting SBFEM-FFT method for representative classes of microstructures, including a short-fibre reinforced composite, a polyurethane foam and a sand-based binder jet microstructure. The results show that the full-voxel SBFEM discretization provides a robust compromise between accuracy and numerical efficiency for both finite-contrast and porous problems. In particular, it remains stable in situations where reduced-integration hexahedral elements feature hourglass instabilities, while avoiding the excessive stiffness typical of fully integrated voxel FEM.

The manuscript is structured as follows. In Section~\ref{sec:SBFEM}, we recall periodic homogenization of linear elastic materials and discretize the cell problem by the SBFEM on a regular grid. Section~\ref{sec:FFT} is devoted to the SBFEM-FFT preconditioner, its spectral estimate and its matrix-free implementation. In Section~\ref{sec:Experiments}, we assess the SBFEM-FFT method for three microstructures and compare it with established voxel discretizations. Section~\ref{sec:conclusion} summarizes our findings.

\section{Discretizing homogenization problems on voxel grids with the SBFEM}\label{sec:SBFEM}
\subsection{Periodic homogenization of linear elastic materials}\label{sec:PeriodicHomogenization}

In this section, we derive the SBFEM discretization of the periodic cell problem on a voxel grid, on which the FFT-based solver of Section~\ref{sec:FFT} operates. To this end, we construct the SBFEM on a reference voxel and condense the radial displacements. Moreover, we evaluate the effective stiffness exactly and show that the SBFEM is softer than fully integrated hexahedral elements.

First, we summarize the key equations of periodic homogenization of linear elastic materials and refer to the literature~\cite{michel1999effective,milton2002theory} for further details.

We consider a periodic and linear elastic homogenization problem posed on the three-dimensional microscopic unit cell
\begin{equation}\label{eq:unitcell}
    Y = [0,\ell_1] \times [0,\ell_2] \times [0,\ell_3]
\end{equation}
with edge lengths $\ell_i$ ($i=1,2,3$), cell volume $|Y| := \ell_1\,\ell_2\,\ell_3$ and cell average $\langle \cdot \rangle_Y := |Y|^{-1}\int_Y \cdot  \diff V$. For each microscopic point $\fx \in Y$, a local material stiffness tensor $\ffC(\fx)\in\Sym{\Sym{3}}$ is given, which we assume to be uniformly bounded via the inequalities
\begin{equation}\label{eq:stiffnessBounds}
    C_-\norm{\feps}^2\leq \feps:\ffC(\fx):\feps \leq C_+\norm{\feps}^2 \quad\text{for all } \feps\in\Sym{3}
\end{equation}
and constants $C_-\in\R_{\geq0}, C_+\in\R_{>0}$.
In linear elasticity, the stress field
\begin{equation}
    \fsigma(\fx) = \ffC(\fx) : \feps(\fx), \quad \fx \in Y
\end{equation}
is computed from the local strain tensor $\feps(\fx)\in\Sym{3}$ at every point $\fx \in Y$.
The strain tensor
\begin{equation}
    \feps(\fx) = \bfeps+\fsymgrad \fu(\fx)\quad\text{for } \fx\in Y
\end{equation}
comprises a homogeneous part, the effective strain $\bfeps$, which we prescribe in a strain-based loading, and a fluctuating part, the symmetrized gradient of the displacement fluctuation field $\fu$ that is an element of the first-order Sobolev space \mbox{$H_{\#}^{1}(Y;\R^3)$} of periodic fields with vanishing mean.\\
In periodic linear elastic homogenization problems, we seek to find a displacement fluctuation field \mbox{$\fu\in H_{\#}^{1}(Y;\R^3)$} that solves the (quasi-)static equilibrium without microscopic body forces
\begin{equation}\label{eq:strongForm}
 \div\left( \ffC(\fx): \left(\bfeps+\fsymgrad \fu(\fx)\right)\right) = \fzero
\end{equation}
for a given local material stiffness tensor field $\ffC$ and a prescribed strain loading $\bfeps$ on the cell $Y$. In weak form, the balance of linear momentum~\eqref{eq:strongForm} requires the displacement fluctuation field \mbox{$\fu\in H_{\#}^{1}(Y;\R^3)$} to satisfy the variational equation
\begin{equation}\label{eq:weakForm}
   \left\langle \fsymgrad \fv : \ffC : \left(\bfeps+\fsymgrad \fu\right) \right\rangle_Y = 0 \quad \text{for all } \fv\in H_{\#}^{1}(Y;\R^3),
\end{equation}
which serves as the starting point for the discretization in Section~\ref{sec:SBFEMdisc}.
By solving the weak form~\eqref{eq:weakForm} for six linearly independent load cases, we may determine the components of the effective material stiffness $\ffC^\text{eff}$, which is defined via the average stress
\begin{equation}\label{eq:effectiveStiffness}
   \ffC^\text{eff} : \bfeps := \left\langle \ffC : \left(\bfeps+\fsymgrad \fu\right) \right\rangle_Y,
\end{equation}
where $\fu$ denotes the solution of the weak form~\eqref{eq:weakForm} for the effective strain $\bfeps$.
Since the stiffness tensor field $\ffC$ is symmetric, the weak form~\eqref{eq:weakForm} is the Euler-Lagrange equation~\cite{blanc2023homogenization,zohdi2008introduction} of the convex elastic energy
\begin{equation}\label{eq:energy}
   \mathcal{E}\left(\fu \right) := \frac{1}{2} \left\langle \left(\bfeps+\fsymgrad \fu\right)  : \ffC: \left(\bfeps+\fsymgrad \fu\right) \right\rangle_Y,
\end{equation}
i.e., its solution minimizes the energy~\eqref{eq:energy} on the space $H_{\#}^{1}(Y;\R^3)$. Testing the weak form~\eqref{eq:weakForm} with the solution $\fu$ itself shows that the minimum value equals $\bfeps:\ffC^\text{eff}:\bfeps/2$, which is known as the Hill-Mandel condition~\cite{hill1963elastic}.
Instead of the full effective strain tensor $\bfeps$, mixed boundary conditions may be prescribed, which fix some components of the effective strain and the complementary components of the effective stress. The extension of the framework discussed below to mixed boundary conditions is straightforward~\cite{kabel2016mixed}, and we use uniaxial strain loadings in the computational investigations of Section~\ref{sec:Experiments}.

\subsection{Discretization via the scaled boundary finite element method}\label{sec:SBFEMdisc}
\begin{figure}
    \centering
    \includegraphics{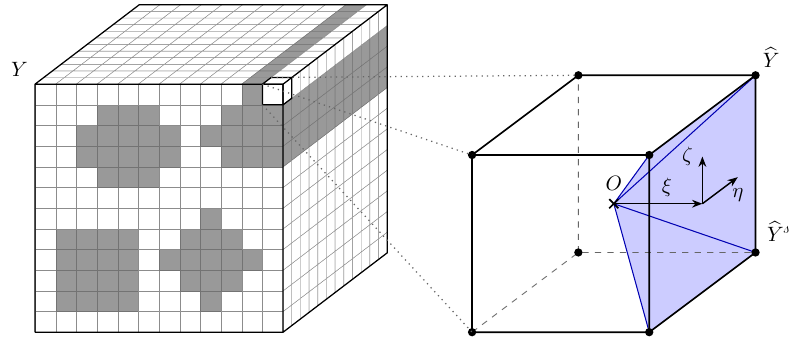}
    \caption{Voxel discretization of a fibre-reinforced microstructure (left) and the reference voxel $\widehat{Y}$ with scaling centre $O$, decomposed into six pyramid-shaped sectors, one of which, the sector $\widehat{Y}^{\sector}$, is shaded together with the local coordinates $(\xi,\eta,\zeta)$ (right). Each voxel $Y^e$ is a translate of the reference voxel.}
    \label{fig:sbsys}
\end{figure}
This section comprises a derivation of the scaled boundary finite element method (SBFEM)~\cite{SongW-97,Wolf-2003,Song-2018}, specialized to three-dimensional periodic homogenization on a regular grid. For an overview of polygonal SBFEM for two-dimensional homogenization, we refer to Section~3 of Talebi et al.~\cite{talebi2019scaled}.

We consider a regular grid of $N_x\times N_y\times N_z$ voxels, which decomposes the cell $Y$ into $n_e := N_x\,N_y\,N_z$ non-overlapping cubic elements $Y^e$ with the edge length $h := \ell_1/N_x = \ell_2/N_y = \ell_3/N_z$. As is common for image-based microstructures, we assume that each voxel is occupied by a single material, i.e., the stiffness tensor field is homogeneous on each element, $\ffC|_{Y^e} \equiv \ffC^e$ for $e=1,\dots,n_e$. Throughout, we use Mandel notation, i.e., we identify strain and stress tensors with vectors in $\R^6$ whose shear components are scaled by $\sqrt{2}$, and the stiffness tensor $\ffC^e$ with a symmetric matrix in $\R^{6\times6}$, and we use the same symbols for both representations. Discrete vectors and matrices, such as nodal displacement vectors and stiffness matrices, are denoted by upright bold symbols.

\parhead{Reference voxel and SBFEM spaces}All elements are translates of the reference voxel
\begin{equation}\label{eq:refVoxel}
    \widehat{Y} := \left[-h/2,h/2\right]^3,
\end{equation}
i.e., the identity $Y^e = \fx_O^e + \widehat{Y}$ holds with the centroid $\fx_O^e$ of the element $Y^e$. Thus, it suffices to construct the SBFEM on the reference voxel. We select its centroid, the origin, as the scaling centre $O$ and divide the reference voxel into six pyramid-shaped sectors $\widehat{Y}^{\sector}$ ($\sector=1,\dots,6$) of identical volume, each of which is the pyramid with one face of the voxel as its base and the scaling centre $O$ as its apex, see Fig.~\ref{fig:sbsys}. The decompositions
\begin{equation}
    Y = \bigcup_{e=1}^{n_e} \left(\fx_O^e + \widehat{Y}\right) \quad \text{with}  \quad \widehat{Y} = \bigcup_{\sector=1}^{6} \widehat{Y}^{\sector}
\end{equation}
hold, where the indices $e$ and $\sector$ specify the element and the sector under consideration, respectively.
In the SBFEM, only the boundary of the voxel, i.e., its six faces, is discretized. While arbitrary surface discretizations, for example, by higher-order triangular and quadrilateral elements~\cite{BirkPAMMdraft}, transition elements~\cite{Gravenkamp-2020} or polygon elements based on Wachspress interpolants~\cite{Natarajan-2017}, are possible, we use one bilinear four-node quadrilateral element per face. On the face of the sector $\widehat{Y}^{\sector}$, the bilinear shape functions $N_i$ ($i=1,\dots,4$) interpolate the coordinates $\widehat{\fx}^{\sector}_i$ of the four corner nodes, and the SBFEM coordinate transform
\begin{equation}
    \widehat{\fx}(\xi,\eta,\zeta) = \xi \sum_{i=1}^{4} N_i(\eta,\zeta) \, \widehat{\fx}^{\sector}_i \label{eq:sbtrans}
\end{equation}
maps the local coordinates $(\xi,\eta,\zeta)\in\Xi\times[-1,1]^2$ of the sector to the points $\widehat{\fx}\in\widehat{Y}^{\sector}$ of the reference voxel. The local coordinates $(\eta,\zeta)$ parametrize the face that bounds the sector, and the radial coordinate \mbox{$\xi \in \Xi := \left[0,1\right]$} scales this face towards the scaling centre, such that the value $\xi=0$ corresponds to the scaling centre and the value $\xi=1$ to the boundary.

The SBFEM combines the interpolation on the boundary with radial displacement functions, which are determined analytically. More precisely, we consider the semi-analytical reference space
\begin{equation}\label{eq:VsemiEl}
\widehat{V}_{\text{semi}} := \left\{ \fd \;\middle|\; \fd(\widehat{\fx}(\xi,\eta,\zeta)) = {\mathbf{N}}(\eta,\zeta)\,{\mathbf{L}^\sector}\, {\mathbf{\displacement}_r}(\xi) \text{ on } \widehat{Y}^{\sector},\ \sector=1,\dots,6, \text{ with } {\mathbf{\displacement}_r}\in H^1\big(\Xi;\R^{24}\big) \right\}.
\end{equation}
Here, the matrix-valued function ${\mathbf{N}}:[-1,1]^2\rightarrow\R^{3\times12}$ collects the shape functions of a face, the matrix ${\mathbf{L}^\sector}\in\R^{12\times24}$ extracts the twelve degrees of freedom of the face bounding the sector $\widehat{Y}^{\sector}$ from the $24$ degrees of freedom of the element, and the vector-valued function ${\mathbf{\displacement}_r}$ comprises the radial displacements of the eight corner nodes. Its value ${\mathbf{\displacement}_b} := {\mathbf{\displacement}_r}(1)\in\R^{24}$ at the boundary collects the nodal displacements of the voxel.

Each function of the reference space~\eqref{eq:VsemiEl} is continuous on the reference voxel, and on each face it coincides with the bilinear interpolant of the nodal displacements at the four corners of that face. Adjacent elements share these corner nodes. Hence, a field $\fu$ whose translated restrictions $\fu(\fx_O^e+\,\cdot\,)$ to the reference voxel lie in the reference space~\eqref{eq:VsemiEl} and whose nodal displacements coincide at shared nodes is continuous across the element faces. Being piecewise of class $H^1$ and continuous, such a field is of class $H^1$~\cite{brenner2008mathematical}. Identifying the nodes on opposite faces of the cell $Y$ accounts for periodicity, and we define the global semi-analytical SBFEM space
\begin{equation}\label{eq:VSemiGlobal}
   V_{\text{semi}} := \left\{ \fu \in H_{\#}^{1}\big(Y;\R^{3}\big) \,\, \middle| \,\, \fu\left(\fx_O^e+\,\cdot\,\right)\big|_{\widehat{Y}}\in \widehat{V}_{\text{semi}} \text{ for } e=1,\dots,n_e\right\}.
\end{equation}
The Galerkin discretization of the weak form~\eqref{eq:weakForm} on the space~\eqref{eq:VSemiGlobal} seeks a displacement fluctuation field $\fu\in V_{\text{semi}}$ that satisfies the variational equation
\begin{equation}\label{eq:discretizedEnergy}
   \left\langle \fsymgrad \fv : \ffC : \left(\bfeps+\fsymgrad \fu\right) \right\rangle_Y = 0 \quad \text{for all } \fv\in V_{\text{semi}}.
\end{equation}
As the inclusion $V_{\text{semi}}\subseteq H_{\#}^{1}(Y;\R^3)$ holds, the SBFEM provides a conforming discretization of the weak form~\eqref{eq:weakForm}.

\parhead{Element-wise solution of the radial equation}The space~\eqref{eq:VSemiGlobal} is infinite-dimensional, since the radial displacements range over the function space $H^1\big(\Xi;\R^{24}\big)$, and thus may not be used directly in a numerical scheme. To arrive at a finite-dimensional problem, we consider two classes of test fields in the Galerkin equation~\eqref{eq:discretizedEnergy}. The first class comprises the test fields $\fv$ in the space~\eqref{eq:VSemiGlobal} that vanish outside of a single element $Y^e$, such that their nodal displacements vanish. Since the stiffness $\ffC^e$ is homogeneous on the element and the test field $\fv$ vanishes on the boundary $\partial Y^e$, the divergence theorem implies the identity
\begin{equation}\label{eq:effectiveStrainDropsOut}
   \int_{Y^e} \fsymgrad \fv : \ffC^e : \bfeps  \diff V = \left(\ffC^e : \bfeps\right) : \int_{\partial Y^e} \operatorname{sym}\left(\fv\otimes\fn\right) \diff A = 0
\end{equation}
with the outward unit normal $\fn$. Thus, the effective strain drops out, and the Galerkin equation~\eqref{eq:discretizedEnergy} for the first class of test fields reduces to the condition
\begin{equation}\label{eq:radialCondition}
   \int_{Y^e} \fsymgrad \fv : \ffC^e : \fsymgrad \fu  \diff V = 0 \quad \text{for all } \fv\in V_{\text{semi}} \text{ vanishing outside of } Y^e
\end{equation}
on each element $Y^e$. In particular, the radial displacements on each element are determined by the nodal displacements of that element alone, independently of the effective strain. In the following, we solve the condition~\eqref{eq:radialCondition} analytically for a generic field $\fd$ in the reference space~\eqref{eq:VsemiEl} with prescribed nodal displacements ${\mathbf{\displacement}_b}$.

We work on the reference voxel $\widehat{Y}$ with the homogeneous stiffness $\ffC^e$ of a fixed element $Y^e$. As the translation $\fx\mapsto\fx-\fx_O^e$ leaves strains unchanged, the quantities derived below depend on the element only via its stiffness $\ffC^e$, and we drop the element index $e$ except for the stiffness $\ffC^e$ and the element stiffness matrix $\mathbf{K}^e$, which enters the assembly. With the SBFEM coordinate transform~\eqref{eq:sbtrans}, the symmetrized gradient in Mandel notation takes the form of the matrix-valued differential operator
\begin{equation}
    \fsymgrad = {\mathbf{b}_1}\frac{\partial}{\partial \xi} + \frac{1}{\xi}\left( {\mathbf{b}_2}\frac{\partial}{\partial \eta}+ {\mathbf{b}_3}\frac{\partial}{\partial \zeta}\right) \label{eq:b1b2b3}
\end{equation}
with the matrix-valued functions ${\mathbf{b}_i}:[-1,1]^2\rightarrow\R^{6\times3}$ ($i=1,2,3$), which are given explicitly in Appendix~\ref{app:coeff}. For a radial displacement ${\mathbf{\displacement}_r}$, the strain vector on the sector $\widehat{Y}^{\sector}$ thus takes the form
\begin{equation}\label{eq:strainSector}
    \feps^\sector({\mathbf{\displacement}_r}) := \fsymgrad\, {\mathbf{N}}\,{\mathbf{L}^\sector}\, {\mathbf{\displacement}_r} = {\mathbf{B}_1^\sector}\, {\mathbf{\displacement}_r}_{,\xi} + \frac{1}{\xi}\,{\mathbf{B}_2^\sector}\, {\mathbf{\displacement}_r}
\end{equation}
involving the matrix-valued functions
\begin{equation}\label{eq:BMatricesSector}
    {\mathbf{B}_1^\sector} := {\mathbf{b}_1}\,{\mathbf{N}}\, {\mathbf{L}^\sector} \qquad \text{and} \qquad {\mathbf{B}_2^\sector} := \left( {\mathbf{b}_2}\,{\mathbf{N}}_{,\eta} + {\mathbf{b}_3}\, {\mathbf{N}}_{,\zeta} \right) {\mathbf{L}^\sector}
\end{equation}
of the local coordinates $(\eta,\zeta)$ with values in $\R^{6\times24}$, where a comma in a subscript followed by a variable denotes the partial derivative with respect to this variable.
The infinitesimal volume element takes the form~\cite[Eq.~(6.73)]{Song-2018}
\begin{equation}\label{eq:volEl} \diff V = \xi^2 \det({\mathbf{J}}) \diff\xi \diff\eta  \diff\zeta,
\end{equation}
where the Jacobian determinant $\det({\mathbf{J}})$ at the boundary is given explicitly in Appendix~\ref{app:coeff}.

Translated to the reference voxel, the test fields in the condition~\eqref{eq:radialCondition} are the fields in the reference space~\eqref{eq:VsemiEl} whose radial displacements vanish at $\xi=1$. Thus, the condition~\eqref{eq:radialCondition} requires the radial displacement ${\mathbf{\displacement}_r}\in H^1\big(\Xi;\R^{24}\big)$ with ${\mathbf{\displacement}_r}(1)={\mathbf{\displacement}_b}$ to satisfy
\begin{equation}\label{eq:firstOrderXi}
 \sum_{\sector=1}^{6} \int_{\widehat{Y}^{\sector}}\T{\feps^\sector({\mathbf{v}_r})}\,\ffC^e\,\feps^\sector({\mathbf{\displacement}_r}) \diff V = 0 \quad\text{for all } {\mathbf{v}_r}\in H^1\big(\Xi;\R^{24}\big) \text{ with } {\mathbf{v}_r}(1)={\mathbf{0}}.
\end{equation}
We introduce the coefficient matrices ${\mathbf{E}_i}\in\R^{24\times24}$ via
\begin{subequations}\label{eq:sbE0E1E2}
\begin{align}
 {\mathbf{E}_{0}}&:=  \sum_{\sector=1}^{6}\int_{-1}^{1}\int_{-1}^{1}\T{{\mathbf{B}_1^\sector}}\,\ffC^e\,{\mathbf{B}_1^\sector}  \,\,\text{det}({\mathbf{J}})  \diff\eta \diff\zeta,\\
 {\mathbf{E}_{1}}&:=  \sum_{\sector=1}^{6}  \int_{-1}^{1}\int_{-1}^{1} \T{{\mathbf{B}_2^\sector}}\,\ffC^e\,{\mathbf{B}_1^\sector} \,\,\text{det}({\mathbf{J}})  \diff\eta \diff\zeta,\\
 {\mathbf{E}_{2}}&:=  \sum_{\sector=1}^{6}\int_{-1}^{1}\int_{-1}^{1} \T{{\mathbf{B}_2^\sector}}\,\ffC^e\,{\mathbf{B}_2^\sector} \,\,\text{det}({\mathbf{J}})  \diff\eta \diff\zeta.
\end{align}
\end{subequations}
For the four-node quadrilateral elements used in this work, closed-form solutions of the integrals~\eqref{eq:sbE0E1E2} are provided in Wolf and Song~\cite{WolfSong1996}.
We insert the volume element~\eqref{eq:volEl} into the condition~\eqref{eq:firstOrderXi} and integrate by parts in the radial coordinate $\xi$ in the terms involving the derivative ${\mathbf{v}_r}_{,\xi}$. The boundary term at $\xi=1$ vanishes because the test field ${\mathbf{v}_r}$ vanishes there, whereas the boundary term at the scaling centre $\xi=0$ vanishes due to the factor $\xi^2$ of the volume element~\eqref{eq:volEl}. As the test field ${\mathbf{v}_r}$ is arbitrary otherwise, the radial displacement ${\mathbf{\displacement}_r}$ solves the scaled boundary finite element equation
\begin{equation}\label{eq:sbfeu}
 {\mathbf{E}_0}\,\xi^2\, {\mathbf{\displacement}_r}(\xi)_{,\xi\xi} + \left( 2{\mathbf{E}_0} - {\mathbf{E}_1} + \T{{\mathbf{E}_1}}\right) \xi \, {\mathbf{\displacement}_r}(\xi)_{,\xi} + \left( \T{{\mathbf{E}_1}} - {\mathbf{E}_2} \right)\, {\mathbf{\displacement}_r}(\xi) = {\mathbf{0}} \quad \text{for all } \xi\in\Xi,
\end{equation}
i.e., an ordinary differential equation on the interval $\Xi$ with values in $\R^{24}$, subject to the boundary condition ${\mathbf{\displacement}_r}(1)={\mathbf{\displacement}_b}$.

Using the internal force-displacement relationship~\cite{saputra2020three}
\begin{equation}
    {\mathbf{q}_r}(\xi) := \xi\left( {\mathbf{E}_0}\,\xi\, {\mathbf{\displacement}_r}(\xi)_{,\xi} + \T{{\mathbf{E}_1}}\, {\mathbf{\displacement}_r}(\xi)\right), \label{eq:qrel}
\end{equation}
we follow the standard SBFEM approach~\cite{Song2004} and rewrite the homogeneous second-order Cauchy-Euler differential equation~\eqref{eq:sbfeu} as a system of first-order differential equations
\begin{equation}
    \xi\begin{bmatrix}
    \xi^{1/2}\, {\mathbf{\displacement}_r}(\xi) \\
    \xi^{-1/2}{\mathbf{q}_r}(\xi)
    \end{bmatrix}_{,\xi} = -{\mathbf{Z}} \begin{bmatrix}
    \xi^{1/2}\, {\mathbf{\displacement}_r}(\xi) \\
    \xi^{-1/2}{\mathbf{q}_r}(\xi)
    \end{bmatrix} \label{eq:fosys}
\end{equation}
with the matrix~\cite[Eq.~(42)]{Song2004}
\begin{equation}
  {\mathbf{Z}} = \begin{bmatrix}
    {\mathbf{Z}_{11}} &{\mathbf{Z}_{12}}\\
    {\mathbf{Z}_{21}} &{\mathbf{Z}_{22}}
  \end{bmatrix}=
  \begin{bmatrix}
    {\mathbf{E}_{0}}^{-1}\T{{\mathbf{E}_{1}}}-\tfrac{1}{2}{\mathbf{I}} &-{\mathbf{E}_{0}}^{-1}\\
     -{\mathbf{E}_{2}}+{\mathbf{E}_{1}}\,{\mathbf{E}_{0}}^{-1}\T{{\mathbf{E}_{1}}} & -{\mathbf{E}_{1}}\,{\mathbf{E}_{0}}^{-1}+\tfrac{1}{2}{\mathbf{I}}
  \end{bmatrix}. \label{eq:zmat}
\end{equation}
The matrix ${\mathbf{Z}}$ is Hamiltonian~\cite{Song2004}, since the relation
\begin{equation}
 {\mathbf{Q}}\,{\mathbf{Z}} = \T{\left({\mathbf{Q}}\,{\mathbf{Z}}\right)} \text{ with } {\mathbf{Q}}=\begin{bmatrix}
    {\mathbf{0}}&{\mathbf{I}}\\
    -{\mathbf{I}}&{\mathbf{0}}\\
 \end{bmatrix}
\end{equation}
holds~\cite{golub2013matrix}, with the $24\times24$-dimensional identity matrix ${\mathbf{I}}$.
The solution of the first-order differential equations~\eqref{eq:fosys} may be obtained as follows. The Hamiltonian matrix ${\mathbf{Z}}\in\R^{48\times48}$ is diagonalizable if and only if, for each of its eigenvalues, the algebraic and the geometric multiplicity coincide, which may only fail for multiple eigenvalues. In the general case, the Hamiltonian matrix ${\mathbf{Z}}$ is therefore decomposed via a Schur decomposition~\cite[Sec.~4]{Song2004}, whereas an eigenvalue decomposition is sufficient if the Hamiltonian matrix is diagonalizable~\cite[Sec.~3]{Song2004}. For the hexahedral elements of the work at hand, multiple eigenvalues occur, but we observe numerically that their algebraic and geometric multiplicities coincide. Therefore, we rely on the eigenvalue decomposition, which is simpler and computationally more efficient than the Schur decomposition, and decompose the Hamiltonian matrix ${\mathbf{Z}}$ in the form
\begin{equation}
    {\mathbf{Z}} = {\mathbf{V}}\,{\mathbf{S}}\,{\mathbf{V}}^{-1}, \label{eq:ZSchur}
\end{equation}
with the diagonal matrix ${\mathbf{S}}$ containing the eigenvalues and the invertible matrix ${\mathbf{V}}$ containing the corresponding eigenvectors as columns. Since the Hamiltonian matrix ${\mathbf{Z}}$ is not normal, eigenvectors corresponding to different eigenvalues are in general not orthogonal, such that the inverse ${\mathbf{V}}^{-1}$ differs from the conjugate transpose. The matrices are decomposed into blocks
\begin{equation} \label{eq:VandS}
  {\mathbf{V}} = \begin{bmatrix}
    {\mathbf{V}_{11}} &{\mathbf{V}_{12}}\\
    {\mathbf{V}_{21}} &{\mathbf{V}_{22}}
  \end{bmatrix}, \qquad
  {\mathbf{S}} = \begin{bmatrix}
    {\mathbf{S}_{11}} &{\mathbf{0}}\\
    {\mathbf{0}} &{\mathbf{S}_{22}}
  \end{bmatrix},
\end{equation}
where the blocks in ${\mathbf{S}}$ are sorted in ascending order with respect to the real part of the corresponding eigenvalue. Since the eigenvalues of a Hamiltonian matrix occur in pairs $\pm \lambda$~\cite{Song-2018,saputra2020three}, the diagonal block ${\mathbf{S}_{11}}$ of dimension $24\times24$ comprises the eigenvalues with negative real part, which are relevant for the solution in bounded domains~\cite{Song-2018,saputra2020three}. Eventually, the solutions for the radial displacement and the internal force are obtained as the vectors
\begin{align}
    {\mathbf{\displacement}_r}(\xi) &= {\mathbf{V}_{11}} \xi^{-{\mathbf{S}_{11}}-{\mathbf{I}}/2}{\mathbf{c}}, \label{eq:sold} \\
    {\mathbf{q}_r}(\xi) &= {\mathbf{V}_{21}}\xi^{-{\mathbf{S}_{11}}+{\mathbf{I}}/2}{\mathbf{c}}. \label{eq:solq}
\end{align}

Here, for a diagonal matrix ${\mathbf{M}}$ of dimension $24\times24$ with complex diagonal entries $M_1,\dots,M_{24}$, we use the abbreviation $\xi^{{\mathbf{M}}} := \operatorname{diag}(\xi^{M_1},\dots,\xi^{M_{24}})$.

Evaluating the general solution~\eqref{eq:sold} at the boundary ($\xi=1$) yields the constant
\begin{equation}\label{eq:constant}
{\mathbf{c}} =  {\mathbf{V}_{11}}^{-1} {\mathbf{\displacement}_b},
\end{equation}
which expresses the integration constants in terms of the nodal displacements.

\parhead{Condensed spaces and element stiffness matrix}Inserting this expression into the general solution~\eqref{eq:sold} provides the explicit representation
\begin{equation}\label{eq:radialExplicit}
    {\mathbf{\displacement}_r}(\xi) = {\mathbf{V}_{11}}\, \xi^{-{\mathbf{S}_{11}}-{\mathbf{I}}/2}\, {\mathbf{V}_{11}}^{-1}{\mathbf{\displacement}_b} \quad \text{for } \xi\in\Xi
\end{equation}
of the radial displacement in terms of the nodal displacements. By the condition~\eqref{eq:radialCondition}, the solution $\fu$ of the Galerkin equation~\eqref{eq:discretizedEnergy} thus lies in the finite-dimensional global space
\begin{equation}\label{eq:VCGlobal}
    V_{\ffC} := \left\{ \fu \in H_{\#}^{1}\big(Y;\R^{3}\big)\,\, \middle| \,\, \fu\left(\fx_O^e+\,\cdot\,\right)\big|_{\widehat{Y}}\in \widehat{V}_{\ffC^e} \text{ for } e=1,\dots,n_e\right\}
\end{equation}
involving the reference spaces
\begin{equation}\label{eq:VCElement}
\widehat{V}_{\ffC^e} := \left\{ \fd\in \widehat{V}_{\text{semi}} \;\middle|\; {\mathbf{\displacement}_r} \text{ is of the form~\eqref{eq:radialExplicit}} \right\}
\end{equation}
of real-valued fields. The semi-analytical space~\eqref{eq:VSemiGlobal} is material-independent but infinite-dimensional, whereas the condensed space~\eqref{eq:VCGlobal} is finite-dimensional but material-dependent, as indicated by the subscript $\ffC$, and the hierarchy $V_{\ffC}\subseteq V_{\text{semi}} \subseteq  H_{\#}^{1}(Y;\R^{3})$ holds. The reference space~\eqref{eq:VCElement} depends on the element only via its stiffness $\ffC^e$. In particular, if the stiffness field $\ffC$ takes only the $n_m$ values $\ffC_1,\dots,\ffC_{n_m}$, at most $n_m$ different reference spaces $\widehat{V}_{\ffC_1},\dots,\widehat{V}_{\ffC_{n_m}}$ arise, and on each element $Y^e$, every field in the condensed space~\eqref{eq:VCGlobal} is the translate of a field in one of these reference spaces.

Inserting the general solutions for the displacement~\eqref{eq:sold} and the force~\eqref{eq:solq} at the boundary $\xi=1$ relates the nodal element force vector ${\mathbf{q}_b}:= {\mathbf{q}_r}(1)$ to the nodal displacements via
\begin{equation}\label{eq:qKd}
  {\mathbf{q}_b} =  {\mathbf{K}^e} \, {\mathbf{\displacement}_b}
\end{equation}
with the element stiffness matrix
\begin{equation}\label{eq:elementStiffnessMatrix}
  {\mathbf{K}^e} := {\mathbf{V}_{21}}\, {\mathbf{V}_{11}}^{-1}.
\end{equation}
The element stiffness matrix ${\mathbf{K}^e}\in\R^{24\times24}$ is symmetric and positive semi-definite, as derived in Chapter~6.12 of Song's book~\cite{Song-2018}, and it is identical for all elements of the same material, such that we compute it once for each material in preprocessing. For pore voxels, i.e., for a vanishing stiffness $\ffC^e=0$, the matrix ${\mathbf{E}_0}$ is singular and the Hamiltonian matrix~\eqref{eq:zmat} is not defined. As the elastic energy of every field vanishes on such voxels, we set ${\mathbf{K}^e} := {\mathbf{0}}$.

From a finite element perspective, the matrix ${\mathbf{K}^e}$ results from a static condensation with the radial displacement ${\mathbf{\displacement}_r}$ in the role of the interior unknowns, i.e., it is the Schur complement with respect to the radial displacement. In contrast to the static condensation of interior degrees of freedom in standard finite element methods, where the interior space is finite-dimensional and the condensation is carried out by Gaussian elimination, the interior space $H^1\big(\Xi;\R^{24}\big)$ is infinite-dimensional, and the condensation is carried out analytically via the scaled boundary finite element equation~\eqref{eq:sbfeu}. In particular, the condensation preserves the element energy. More precisely, the identity
\begin{equation}\label{eq:condensedBilinear}
  \int_{\widehat{Y}} \fsymgrad \fw : \ffC^e : \fsymgrad \fd  \diff V = \T{{\mathbf{w}_b}}\,{\mathbf{K}^e}\,{\mathbf{\displacement}_b}
\end{equation}
holds for all fields $\fw$ in the reference space~\eqref{eq:VsemiEl} and $\fd$ in the reference space~\eqref{eq:VCElement} with nodal displacements ${\mathbf{w}_b},{\mathbf{\displacement}_b}\in\R^{24}$. Indeed, we insert the volume element~\eqref{eq:volEl} and integrate by parts in the radial coordinate $\xi$ in the terms involving the derivative ${\mathbf{w}_r}_{,\xi}$, as in the derivation of the scaled boundary finite element equation~\eqref{eq:sbfeu}. The remaining integral vanishes by the scaled boundary finite element equation~\eqref{eq:sbfeu}, the boundary term at the scaling centre $\xi=0$ vanishes by the representation~\eqref{eq:radialExplicit}, and the boundary term at $\xi=1$ equals $\T{{\mathbf{w}_b}}\,{\mathbf{q}_b}$, which coincides with the right-hand side of the identity~\eqref{eq:condensedBilinear} by the relation~\eqref{eq:qKd}.

\parhead{Assembly of the global system}
With the condensed space~\eqref{eq:VCGlobal}, the element stiffness matrix~\eqref{eq:elementStiffnessMatrix} and the identity~\eqref{eq:condensedBilinear}, the Galerkin equation~\eqref{eq:discretizedEnergy} takes the form
\begin{equation}\label{eq:VectorMatrixFirstOrder}
  \sum_{e=1}^{n_e} \T{{\mathbf{v}^e_b}}\,{\mathbf{K}^e} \left({\mathbf{u}^e_b} + {\mathbf{T}}\,\bfeps\right) = 0 \quad \text{for all test fields } \fv\in V_{\ffC}
\end{equation}
where the vectors ${\mathbf{u}^e_b},{\mathbf{v}^e_b}\in\R^{24}$ denote the nodal displacements of the fields $\fu$ and $\fv$ on the element $Y^e$, and the vector
\begin{equation}\label{eq:linearMapping}
    {\mathbf{T}} \,\bfeps := \left( \bar{\feps}\left(\widehat{\fx}_i-\widehat{\fx}_1 \right)\right)_{i=1}^8 \in \R^{24}
\end{equation}
collects the nodal displacements induced by the effective strain $\bfeps$ at the corner nodes $\widehat{\fx}_i$ ($i=1,\dots,8$) of the reference voxel. Two arguments lead to this form. First, it suffices to consider the second class of test fields, that is, the test fields in the condensed space $V_{\ffC}$. Every field in the semi-analytical space~\eqref{eq:VSemiGlobal} is the sum of a field in the space $V_{\ffC}$ with the same nodal displacements and a field with vanishing nodal displacements. The latter vanishes on all element boundaries and is thus a sum of test fields of the first class, for which the Galerkin equation~\eqref{eq:discretizedEnergy} reduces to the condition~\eqref{eq:radialCondition}, satisfied by construction of the condensed space. Second, the effective strain enters through the affine displacement field $\fx\mapsto\bfeps\,(\fx-\fx_R)$ with a global reference point $\fx_R$. On each element $Y^e$, the translated restriction of this field lies in the reference space~\eqref{eq:VsemiEl} and satisfies the condition~\eqref{eq:radialCondition} due to the identity~\eqref{eq:effectiveStrainDropsOut}, that is, it lies in the reference space~\eqref{eq:VCElement}, such that the identity~\eqref{eq:condensedBilinear} applies to it. At the corner nodes $\fx_O^e+\widehat{\fx}_i$ of the element $Y^e$, its nodal displacements differ from the vector~\eqref{eq:linearMapping} only by the rigid translation $\bfeps\,(\fx_O^e+\widehat{\fx}_1-\fx_R)$, which lies in the kernel of the element stiffness matrix ${\mathbf{K}^e}$. The vector~\eqref{eq:linearMapping} is identical for all elements of the regular grid, as shown in Fig.~\ref{fig:elementUwithEps}.

The extraction matrices \mbox{${\mathbf{\Omega}^e}\in\R^{3n_n\times24}$} link the element degrees of freedom to the global degrees of freedom via the localization relation
\begin{equation}\label{eq:localization}
    {\mathbf{u}^e_b}=\T{{\mathbf{\Omega}^e}}{\mathbf{u}},
\end{equation}
where the vector ${\mathbf{u}}\in\R^{3n_n}$ comprises the nodal displacement fluctuations and the scalar $n_n$ denotes the number of global nodes. The periodic boundary conditions are accounted for by the global node numbering encoded in the extraction matrices ${\mathbf{\Omega}^e}$, and the mean-free condition is imposed explicitly in our implementation, as described in Section~\ref{sec:implementation}. With the global stiffness matrix
\begin{equation}\label{eq:globalStiffnessMatrix}
   {\mathbf{K}} := \sum_{e=1}^{n_e} {\mathbf{\Omega}^e}\,{\mathbf{K}^e}\,\T{{\mathbf{\Omega}^e}}
\end{equation}
and the global vector
\begin{equation}
{\mathbf{b}} :=  -\sum_{e=1}^{n_e} {\mathbf{\Omega}^e}\,{\mathbf{K}^e}\, {\mathbf{T}}\,\bfeps,
\end{equation}
which accounts for the effective strain, the variational equation~\eqref{eq:VectorMatrixFirstOrder} is equivalent to the linear system
\begin{equation}\label{eq:linearSystem}
   {\mathbf{K}} \,{\mathbf{u}} = {\mathbf{b}}.
\end{equation}
It states that the nodal element forces are in equilibrium at every node. In Section~\ref{sec:FFT}, we discuss how to solve the linear system~\eqref{eq:linearSystem} efficiently using FFT-based solvers without assembling the system matrix ${\mathbf{K}}$.

\subsection{Effective stiffness of the SBFEM: exact evaluation and upper bound}\label{sec:stressAverage}

In analogy to Eq.~\eqref{eq:effectiveStiffness}, we introduce the effective stiffness $\ffC^\text{eff}_{\text{SBFEM},h}$ of the SBFEM discretization via
\begin{equation}\label{eq:effStiffSBFEM}
    \ffC^\text{eff}_{\text{SBFEM},h}:\bfeps := \left\langle \ffC:\left(\bfeps+\fsymgrad \fu\right)\right\rangle_Y \quad \text{for all } \bfeps\in\Sym{3},
\end{equation}
where $\fu\in V_{\text{semi}}$ denotes the solution of the Galerkin equation~\eqref{eq:discretizedEnergy} for the effective strain $\bfeps$. To compute this average stress, we use the exact expression
   \begin{equation}\label{eq:normSigma}
    \left\langle \fsigma({\mathbf{u}})\right\rangle_Y  =  \sum_{e=1}^{n_e} \frac{1}{n_e}\left\langle \fsigma({\mathbf{u}^e_b} )\right\rangle_{Y^e},
\end{equation}
leveraging the property that the hexahedral elements have identical volumes due to the regular grid.
The average stress vector on each hexahedral element may be expressed exactly in the form
\begin{equation}\label{eq:stressAverage}
   \left\langle \fsigma({\mathbf{u}^e_b} )\right\rangle_{Y^e}  = \ffC^e\,\averageStrainMatrix\, \left({\mathbf{u}^e_b} +  {\mathbf{T}}\, \bfeps\right)
\end{equation}
with the matrix $\averageStrainMatrix$ that maps the nodal displacements of the hexahedral element to the average strain on the hexahedral element and the material stiffness matrix in Mandel notation $\ffC^e$. We compute the matrix $\averageStrainMatrix$ based on the expression
\begin{equation}\label{eq:averageStrainMatrix}
   \averageStrainMatrix =  \frac{1}{|\widehat{Y}|} \left(- {\overline{\mathbf{B}}_1} \, {\mathbf{V}_{11}} \left( {\mathbf{S}_{11}}+ \tfrac{1}{2} {\mathbf{I}} \right) + {\overline{\mathbf{B}}_2} \, {\mathbf{V}_{11}}\right) \left(- {\mathbf{S}_{11}} + \tfrac{3}{2} {\mathbf{I}}\right)^{-1} {\mathbf{V}_{11}}^{-1},
\end{equation}
which follows from Cui et al.~\cite[Eq.(32),(41),(42)]{cui2024scaled}, using the matrices ${\mathbf{S}_{11}},{\mathbf{V}_{11}}$ introduced in Eq.~\eqref{eq:VandS} and the volume $|\widehat{Y}|$ of the reference voxel. Such an exact volume average, derived from the SBFEM governing equations, was used for the homogenization of two-dimensional SBFEM polygons by Wang et al.~\cite{wang2017efficient}.
The matrices ${\overline{\mathbf{B}}_1}$ and ${\overline{\mathbf{B}}_2}$ are defined in the form
\begin{equation}
   {\overline{\mathbf{B}}_i} =  \sum_{\sector=1}^{6} \int_{-1}^1 \int_{-1}^1 {\mathbf{B}_i^\sector}(\eta,\zeta) \,\,\text{det}({\mathbf{J}})  \diff\eta  \diff\zeta  \quad\text{for } i=1,2
\end{equation}
based on the sector matrices~\eqref{eq:BMatricesSector}.
In particular, the averaging operator $ \averageStrainMatrix$ satisfies the affine reproduction identity
\begin{equation}\label{eq:affine}
     \averageStrainMatrix \, {\mathbf{T}} \,\bfeps =  \bfeps,
\end{equation}
i.e., the averaging operator $\averageStrainMatrix$ exactly reproduces homogeneous strain fields from affine displacement data, which is a prerequisite for the convergence of the effective stress. Indeed, the affine field $\widehat{\fx}\mapsto\bar{\feps}\,(\widehat{\fx}-\widehat{\fx}_1)$ lies in the reference space~\eqref{eq:VCElement}, see Section~\ref{sec:SBFEMdisc}, and its nodal displacements are given by the vector~\eqref{eq:linearMapping}. A field in the reference space~\eqref{eq:VCElement} is determined by its nodal displacements via the representation~\eqref{eq:radialExplicit}. Thus, the SBFEM field with the nodal displacements ${\mathbf{T}}\,\bfeps$ coincides with the affine field, whose average strain equals $\bfeps$.

We note that in our regular grid setting, it is sufficient to precompute and cache the matrix $\averageStrainMatrix$ once per material, as it is identical for all hexahedral elements of the same material. Thus, the exact average stress~\eqref{eq:stressAverage} costs one matrix-vector product per element, as does the stress at the scaling centre, by which Saputra et al.~\cite{saputra2020three} and Talebi et al.~\cite{talebi2019scaled} approximate the element average. We work with the exact average stress~\eqref{eq:stressAverage}, as Wang et al.~\cite{wang2017efficient}, because it is the average stress of the Galerkin solution and thus provides a consistent stopping criterion~\eqref{eq:stoppingCriterion} for the FFT-based solvers of Section~\ref{sec:FFT}.
For the SBFEM, the error in the effective stress decreases linearly with the grid spacing $h$, as shown in Section~\ref{sec:Experiments}. This rate is twice the rate of the local strain and stress fields in the $L^2$-norm, which is limited by the material interfaces. The phenomenon is known as superconvergence of the effective properties~\cite{ye2023convergence,schneider2023superconvergence} and is caused by error cancellations upon averaging.

We close this section by comparing the effective stiffness~\eqref{eq:effStiffSBFEM} with the effective stiffness of fully integrated trilinear hexahedral elements (HEX8) on the same voxel grid. For a voxel-wise homogeneous stiffness, the $2\times2\times2$ Gauss quadrature integrates the HEX8 element stiffness matrix exactly, such that HEX8 elements are the Galerkin discretization of the weak form~\eqref{eq:weakForm} on the space $V_{\text{HEX8}}\subseteq H_{\#}^{1}(Y;\R^3)$ of continuous fields that are trilinear on each element. As the voxels are cubes with the scaling centre at the centroid, the inclusion of the space $V_{\text{HEX8}}$ in the semi-analytical space~\eqref{eq:VSemiGlobal} holds.

Indeed, on the reference voxel, the trilinear fields are spanned by the monomials $m(\widehat{\fx}) = \widehat{x}_1^{\alpha_1}\,\widehat{x}_2^{\alpha_2}\,\widehat{x}_3^{\alpha_3}$ with exponents $\alpha_j\in\{0,1\}$. Each monomial is homogeneous of degree $k := \alpha_1+\alpha_2+\alpha_3$ with respect to the scaling centre, i.e., the relation $m(\xi\,\widehat{\fx}) = \xi^k\,m(\widehat{\fx})$ holds. On each face of the voxel, one coordinate is constant, such that the monomial is bilinear in the local coordinates $(\eta,\zeta)$ and coincides with its bilinear interpolant. Thus, the representation
\begin{equation}
    m\big(\widehat{\fx}(\xi,\eta,\zeta)\big) = \xi^k \sum_{i=1}^{4} N_i(\eta,\zeta)\, m\big(\widehat{\fx}^{\sector}_i\big) \quad \text{on } \widehat{Y}^{\sector}
\end{equation}
holds for every sector, i.e., each monomial lies in the reference space~\eqref{eq:VsemiEl} with a radial displacement proportional to $\xi^k$. As translations preserve trilinearity, the inclusion $V_{\text{HEX8}}\subseteq V_{\text{semi}}$ follows.

For a Galerkin discretization on a subspace $W_h\subseteq H_{\#}^{1}(Y;\R^3)$, the discrete solution minimizes the energy~\eqref{eq:energy} on the subspace $W_h$. As for the continuous problem, the minimum value equals $\bfeps:\ffC^\text{eff}_{W_h}:\bfeps/2$ with the effective stiffness $\ffC^\text{eff}_{W_h}$ defined via the average stress~\eqref{eq:effectiveStiffness} of the discrete solution. Thus, the effective stiffness $\ffC^\text{eff}_{W_h}$ does not increase if the subspace $W_h$ is enlarged and is bounded from below by the exact effective stiffness $\ffC^\text{eff}$. As the effective stiffness~\eqref{eq:effStiffSBFEM} is the effective stiffness $\ffC^\text{eff}_{W_h}$ for the subspace $W_h=V_{\text{semi}}$, we obtain the inequalities
\begin{equation}\label{eq:energyOrdering}
   \bfeps:\ffC^\text{eff}:\bfeps \leq \bfeps:\ffC^\text{eff}_{\text{SBFEM},h}:\bfeps \leq \bfeps:\ffC^\text{eff}_{\text{HEX8},h}:\bfeps \quad \text{for all } \bfeps\in\Sym{3}
\end{equation}
for the effective stiffnesses of the SBFEM and of HEX8 elements on the same voxel grid with edge length $h$. In particular, the SBFEM discretization is softer than fully integrated hexahedral elements at every resolution. Reduced-integration hexahedral elements and the tetrahedron-based discretization~\cite{finel2025tetrahedron} do not fit into this framework of conforming Galerkin discretizations, and no such ordering is available for them.

\section{FFT-based solvers for the SBFEM system}\label{sec:FFT}
\subsection{General framework}\label{sec:FFTIntro}

The SBFEM discretization leads to the linear system~\eqref{eq:linearSystem}. In this section, we solve it by an FFT-based method. To this end, we construct an SBFEM-FFT preconditioner from the homogeneous SBFEM stiffness and bound the condition number of the preconditioned system by the material contrast. Moreover, we describe the matrix-free implementation.

First, we recall how FFT-based methods solve this system: an iterative scheme is combined with a preconditioner that is applied in Fourier space. The convergence of iterative solvers for the linear system~\eqref{eq:linearSystem} is governed by the condition number of the system matrix~\cite{saad2003iterative}. For voxel discretizations, the condition number of the stiffness matrix~\eqref{eq:globalStiffnessMatrix} grows with the resolution and with the material contrast. Preconditioning remedies this deficiency. For a symmetric and positive definite matrix ${\mathbf{K}_0}\in\R^{3n_n\times 3n_n}$, we consider the preconditioner ${\mathbf{P}} := {\mathbf{K}_0}^{-1}$ and the preconditioned system
\begin{equation}\label{eq:preconditionedSystem}
  {\mathbf{P}}\left({\mathbf{K}}\,{\mathbf{u}}-{\mathbf{b}}\right) = {\mathbf{0}},
\end{equation}
which shares its solution with the linear system~\eqref{eq:linearSystem}. For the preconditioned system, the condition number of the symmetric matrix ${\mathbf{P}}^{1/2}\,{\mathbf{K}}\,{\mathbf{P}}^{1/2}$ governs the convergence~\cite{saad2003iterative}. Suppose that the inequalities
\begin{equation}\label{eq:inequ}
    \boxed{C_-\, \T{{\mathbf{u}}}\, {\mathbf{K}_0}\, {\mathbf{u}}
    \ \leq\ \T{{\mathbf{u}}}\, {\mathbf{K}}\, {\mathbf{u}}
    \ \leq\ C_+\, \T{{\mathbf{u}}}\, {\mathbf{K}_0}\, {\mathbf{u}}} \quad \text{for all } {\mathbf{u}}\in\R^{3n_n}
\end{equation}
hold with positive constants $C_\pm$. For the vector ${\mathbf{w}}={\mathbf{K}_0}^{1/2}{\mathbf{u}}$, the inequalities~\eqref{eq:inequ} take the form
\begin{equation}
C_- \norm{{\mathbf{w}}}^2 \leq \T{{\mathbf{w}}} {\mathbf{K}_0}^{-1/2}\, {\mathbf{K}}\, {\mathbf{K}_0}^{-1/2}\,{\mathbf{w}} \leq C_+ \norm{{\mathbf{w}}}^2,
\end{equation}
which bounds the condition number of the preconditioned system matrix via
\begin{equation}\label{eq:materialContrast}
  \text{cond}\left( {\mathbf{P}}^{1/2}\, {\mathbf{K}}\,{\mathbf{P}}^{1/2}\right)\leq \kappa \quad \text{with the material contrast} \quad \kappa := \frac{C_+}{C_-}.
\end{equation}

In FFT-based methods, the preconditioner needs to satisfy the properties \propref{a} and \propref{b} stated in the introduction~\cite{schneider2021review}. Property \propref{a} holds if the matrix ${\mathbf{K}_0}$ has homogeneous coefficients, i.e., if it is assembled from a single element matrix on the regular periodic grid. Then, the matrix ${\mathbf{K}_0}$ corresponds to a matrix-valued Fourier multiplier~\cite{rudin1962fourier}, and its inverse is applied independently for each frequency. Property \propref{b} holds if the constants $C_\pm$ in the inequalities~\eqref{eq:inequ} do not depend on the grid spacing $h$.
For voxel-based finite elements~\cite{schneider2017fft,leuschner2018fourier,ladecky2023optimal}, the matrix ${\mathbf{K}_0}$ is the stiffness matrix of the same discretization for a reference material, for example a scaled identity~\cite{schneider2022voxel}. As the shape functions do not depend on the material, both quadratic forms in the inequalities~\eqref{eq:inequ} evaluate the elastic energy of the same displacement field. Hence, the inequalities~\eqref{eq:inequ} follow directly from the material bounds~\eqref{eq:stiffnessBounds}, applied pointwise. For the SBFEM, in contrast, the shape functions depend on the material, see the condensed space~\eqref{eq:VCGlobal}. Thus, the two quadratic forms in the inequalities~\eqref{eq:inequ} evaluate the energies of two different displacement fields, which merely share their nodal displacements, and the material bounds~\eqref{eq:stiffnessBounds} do not apply directly. The inequalities~\eqref{eq:inequ} for the SBFEM therefore require an argument of their own, which we provide in Section~\ref{sec:Preconditioner}.

The simplest FFT-based solution scheme is the basic scheme~\cite{moulinec1994fast,moulinec1998numerical}, which takes the form
\begin{equation}\label{eq:basicScheme}
    {\mathbf{u}}^{k+1} = {\mathbf{u}}^{k} - s \,\invfourier{\fourier{{\mathbf{P}}} \fourier{{\mathbf{r}}^k}} \quad \text{with} \quad {\mathbf{r}}^k = {\mathbf{K}}\,{\mathbf{u}}^{k} - {\mathbf{b}}
\end{equation}
with the discrete Fourier transform $\mathcal{F}$, the inverse discrete Fourier transform $\mathcal{F}^{-1}$ and the step size $s$. The basic scheme~\eqref{eq:basicScheme} may be interpreted as a gradient descent method~\cite{kabel2014efficient} or as a Richardson iteration~\cite{mishra2016comparative} for the preconditioned system~\eqref{eq:preconditionedSystem}. The material contrast $\kappa$ determines how fast the basic scheme~\eqref{eq:basicScheme} converges. With the step size $s=2/(C_++C_-)$, whose choice we discuss in Section~\ref{sec:implementation}, every iteration reduces the error in the norm induced by the matrix~${\mathbf{K}_0}$ at least by the factor $(\kappa-1)/(\kappa+1)$~\cite{saad2003iterative}. Thus, the number of iterations required to reach a prescribed relative tolerance scales linearly with the material contrast~\eqref{eq:materialContrast} and does not depend on the resolution, provided that property~\propref{b} holds. The convergence may be accelerated by further solution schemes from large-scale optimization, which rely on the same principle of a preconditioner applied locally in Fourier space, such as the Barzilai-Borwein scheme~\cite{schneider2019barzilai}, the linear conjugate gradient method~\cite{zeman2010accelerating} and the nonlinear conjugate gradient method~\cite{schneider2020dynamical}. The linear conjugate gradient method reduces the error in the energy norm at least by the factor $(\sqrt{\kappa}-1)/(\sqrt{\kappa}+1)$ per iteration, up to a constant~\cite{saad2003iterative,golub2013matrix}, such that its iteration count grows only with the square root of the material contrast~\eqref{eq:materialContrast}. The iteration counts of these schemes are likewise observed not to depend on the resolution, provided that property~\propref{b} holds.

\subsection{Constructing an SBFEM-FFT preconditioner}\label{sec:Preconditioner}

In this section, we construct the matrix ${\mathbf{K}_0}$ for the SBFEM and provide the arguments for the inequalities~\eqref{eq:inequ}. The established preconditioner in FEM-FFT~\cite{schneider2017fft} may be rewritten~\cite{schneider2022voxel} such that it relies on a scaled identity material matrix, where the scaling factors depend on the maximum/minimum eigenvalue of the material stiffness matrix. We adapt this idea and carry out the construction of Section~\ref{sec:SBFEMdisc} for an identity material stiffness matrix $\ffC^e={\mathbf{I}}$, which in Mandel notation represents the identity tensor on symmetric tensors. The homogeneous element stiffness matrix
\begin{equation}\label{eq:KIElement}
    {\mathbf{K}_0^e} = {\mathbf{V}_{21}^{0}}\,\big({\mathbf{V}_{11}^{0}}\big)^{-1} \ \in\ \R^{24\times24}
\end{equation}
results with the matrices ${\mathbf{V}_{21}^{0}},{\mathbf{V}_{11}^{0}}$ that follow from eigendecomposition of the respective Hamiltonian matrix.
Since the hexahedral elements of the regular grid are uniform, the homogeneous element stiffness matrix ${\mathbf{K}_0^e}$ is identical for every element $e$ and may be computed once in preprocessing. Assembling yields the homogeneous SBFEM stiffness matrix
\begin{equation}\label{eq:K0Global}
    {\mathbf{K}_0} := \sum_{e=1}^{n_e} {\mathbf{\Omega}^e}\,{\mathbf{K}_0^e}\,\T{{\mathbf{\Omega}^e}} \ \in\ \R^{3n_n\times3n_n},
\end{equation}
which has homogeneous coefficients and thus satisfies property \propref{a}. We apply its inverse locally in Fourier space by multiplying each nodal value with a $3\times3$ matrix-valued Fourier multiplier, using a similar procedure to the one developed for trilinear hexahedral finite elements~\cite{schneider2017fft,leuschner2018fourier,ladecky2023optimal}, see Section~\ref{sec:implementation}.

It remains to provide the arguments for the inequalities~\eqref{eq:inequ}, which rely on a minimum characterization of the element stiffness matrix. As the stiffness $\ffC^e$ is symmetric and positive semi-definite, the condition~\eqref{eq:firstOrderXi} is the first-order optimality condition of a convex quadratic minimization problem in the radial displacement. Thus, the radial displacement~\eqref{eq:radialExplicit} minimizes the element energy among all radial displacements with the same nodal displacements, and the identity~\eqref{eq:condensedBilinear} for $\fw=\fd$ implies
\begin{equation}\label{eq:localdiscretizedEnergy}
    \T{{\mathbf{\displacement}_b}}\,{\mathbf{K}^e}\,{\mathbf{\displacement}_b} = \min_{\substack{{\mathbf{\displacement}_r}\in H^1(\Xi;\R^{24})\\ {\mathbf{\displacement}_r}(1)={\mathbf{\displacement}_b}}} \sum_{\sector=1}^{6}\int_{\widehat{Y}^{\sector}}\T{\feps^\sector({\mathbf{\displacement}_r})}\,\ffC^e\,\feps^\sector({\mathbf{\displacement}_r}) \diff V \quad \text{for all } {\mathbf{\displacement}_b}\in\R^{24}.
\end{equation}
For fixed nodal displacements ${\mathbf{\displacement}_b}\in\R^{24}$, we denote by ${\mathbf{\displacement}_r^{C}}$ and ${\mathbf{\displacement}_r^{I}}$ the minimizers of the problem~\eqref{eq:localdiscretizedEnergy} for the stiffness $\ffC^e$ and for the identity ${\mathbf{I}}$, respectively. Both problems share the admissible set $\{{\mathbf{\displacement}_r}\in H^1(\Xi;\R^{24}) \,|\, {\mathbf{\displacement}_r}(1)={\mathbf{\displacement}_b}\}$, which does not depend on the material. Thus, each minimizer is an admissible competitor for the other problem, and the material bounds~\eqref{eq:stiffnessBounds} imply the inequalities
\begin{align}
    \T{{\mathbf{\displacement}_b}}\,{\mathbf{K}^e}\,{\mathbf{\displacement}_b}
    &\leq \sum_{\sector=1}^6\int_{\widehat{Y}^{\sector}} \T{\feps^\sector({\mathbf{\displacement}_r^{I}})}\,\ffC^e\,\feps^\sector({\mathbf{\displacement}_r^{I}}) \diff V
    \leq C_+ \sum_{\sector=1}^6\int_{\widehat{Y}^{\sector}} \norm{\feps^\sector({\mathbf{\displacement}_r^{I}})}^2 \diff V
    = C_+\, \T{{\mathbf{\displacement}_b}}\,{\mathbf{K}_0^e}\,{\mathbf{\displacement}_b},\\
    C_-\, \T{{\mathbf{\displacement}_b}}\,{\mathbf{K}_0^e}\,{\mathbf{\displacement}_b}
    &\leq C_- \sum_{\sector=1}^6\int_{\widehat{Y}^{\sector}} \norm{\feps^\sector({\mathbf{\displacement}_r^{C}})}^2 \diff V
    \leq \sum_{\sector=1}^6\int_{\widehat{Y}^{\sector}} \T{\feps^\sector({\mathbf{\displacement}_r^{C}})}\,\ffC^e\,\feps^\sector({\mathbf{\displacement}_r^{C}}) \diff V
    = \T{{\mathbf{\displacement}_b}}\,{\mathbf{K}^e}\,{\mathbf{\displacement}_b}.
\end{align}
Hence, the energy bounds
\begin{equation}\label{eq:elementSpectralEquiv}
    C_-\, \T{{\mathbf{\displacement}_b}}\, {\mathbf{K}_0^e}\, {\mathbf{\displacement}_b}
    \ \leq\ \T{{\mathbf{\displacement}_b}}\, {\mathbf{K}^e}\, {\mathbf{\displacement}_b}
    \ \leq\ C_+\, \T{{\mathbf{\displacement}_b}}\, {\mathbf{K}_0^e}\, {\mathbf{\displacement}_b}
\end{equation}
hold for all nodal displacement vectors ${\mathbf{\displacement}_b}\in\R^{24}$. The argument is short, but it hinges on the semi-analytical construction. The space~\eqref{eq:VSemiGlobal} does not depend on the material, which enters only via the radial minimization~\eqref{eq:localdiscretizedEnergy}.

Summing the element-wise bounds~\eqref{eq:elementSpectralEquiv} over all elements and using the localization relation~\eqref{eq:localization} together with the assembly formulas~\eqref{eq:globalStiffnessMatrix} and~\eqref{eq:K0Global} yields the inequalities~\eqref{eq:inequ}. Thus, the bound~\eqref{eq:materialContrast} on the condition number holds for the SBFEM, and it coincides with the bound on the condition number of a preconditioned standard voxel-FEM matrix. For porous materials, the relation \mbox{$C_- = 0$} applies, and the bound~\eqref{eq:materialContrast} on the condition number is formally infinite. Nevertheless, FFT-based solvers behave as if the preconditioned system had a finite effective condition number~\cite{schneider2020lippmann}, at least in the mechanically stable case, i.e., if the interface between the solid and the pores is Lipschitz-continuous and the solid phase admits no infinitesimal rigid body motions other than translations. The iterates remain in a subspace of displacement fields that are elastically extended into the pores, and on this subspace, the material contrast is replaced by an effective contrast, which additionally involves a constant determined by the geometry of the interface~\cite{schneider2020lippmann}. In contrast to the material contrast~\eqref{eq:materialContrast}, this effective contrast depends on the discretization. Some discretizations are unstable for porous microstructures, such as the Moulinec-Suquet discretization~\cite{schneider2016computational} and wavelet or finite-difference discretizations with wide stencils~\cite{dave2026limitations}, whereas others are stable, for instance the staggered grid~\cite{schneider2016computational} and certain finite element variants~\cite{schneider2020lippmann}. We investigate the stability of the SBFEM for a porous microstructure numerically in Section~\ref{sec:polyurethaneFoam}.

\subsection{Implementation}\label{sec:implementation}

\begin{figure}
  \centering
  \includegraphics[width=0.8\textwidth]{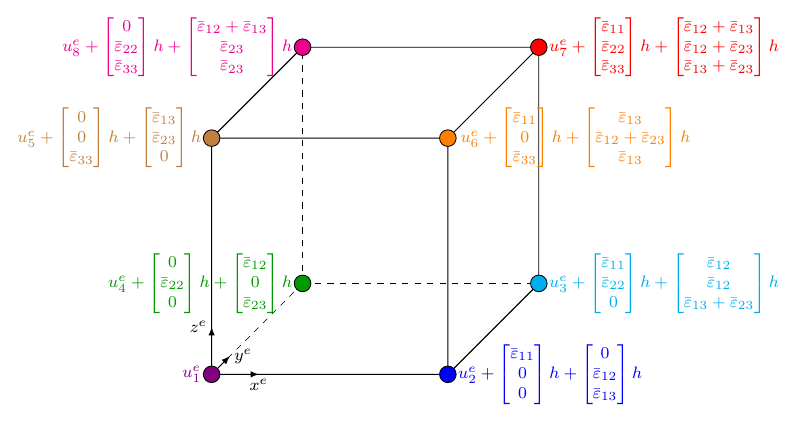}
\caption{Sketch on the displacement computation within a hexahedral element with voxel length $h$ along all edges.}
\label{fig:elementUwithEps}
\end{figure} 

In this section, we discuss the implementation of the basic scheme~\eqref{eq:basicScheme} for the preconditioned SBFEM system~\eqref{eq:preconditionedSystem}.

The SBFEM-FFT preconditioner, derived in the previous section, bounds the eigenvalues of the preconditioned system by the constants $C_\pm$ of the material bounds~\eqref{eq:stiffnessBounds}, see the inequalities~\eqref{eq:inequ}. Based on these bounds, the optimal step size of the basic scheme takes the form 
\begin{equation}
    s = \frac{2}{C_{+}+C_{-}},
    \label{eq:stepsize}
\end{equation}
which results naturally by interpreting the basic scheme as a version of the Richardson iteration~\cite{mishra2016comparative} and using Example 4.1 in Saad's book~\cite{saad2003iterative}.

To implement the basic scheme~\eqref{eq:basicScheme} without assembling the global stiffness matrix~\eqref{eq:globalStiffnessMatrix}, we closely follow the established displacement-based implementations~\cite{leuschner2018fourier,schneider2022voxel,ladecky2023optimal}.
We compute the force vector on the hexahedral element
\begin{equation}\label{eq:resVector}
    {\mathbf{r}^e}({\mathbf{u}^e}) =  {\mathbf{K}^e}\,\left({\mathbf{u}^e}+ {\mathbf{T}}\,\bfeps\right)
\end{equation}
using the element stiffness matrix \mbox{${\mathbf{K}^e}\in\R^{24\times24}$}, the element displacement fluctuation vector \mbox{${\mathbf{u}}^e\in\R^{24}$} and the position-independent vector~\eqref{eq:linearMapping}, see Fig.~\ref{fig:elementUwithEps}. We note that, although the element stiffness matrix~\eqref{eq:elementStiffnessMatrix} is exactly symmetric~\cite{Song-2018}, we additionally symmetrize it numerically to remove residual round-off asymmetry introduced by floating-point arithmetic during the eigendecomposition. As the element stiffness matrix is computed once for each material in preprocessing, this symmetrization is carried out once and for all.
After computing the element force vector~\eqref{eq:resVector} on all hexahedral elements, the global nodal force vector 
\begin{equation}\label{eq:nodalRes}
   {\mathbf{r}}({\mathbf{u}}) =  \sum_{e=1}^{n_e} {\mathbf{\Omega}^e} \, \, {\mathbf{r}^e}(\T{{\mathbf{\Omega}^e}}\,{\mathbf{u}})
\end{equation}
is assembled using the extraction matrices \mbox{${\mathbf{\Omega}^e}\in\R^{3n_n\times24}$}.

In Section~\ref{sec:Preconditioner}, we derived the inverse of the homogeneous SBFEM stiffness matrix, i.e., the SBFEM Green's operator, as a suitable SBFEM-FFT preconditioner, such that the relation ${\mathbf{P}}={\mathbf{K}_0}^{-1}$ holds.
We compute the matrix that represents the action of the SBFEM-FFT preconditioner using a similar procedure as for the FEM Green's operator in FFT-solvers~\cite{schneider2017fft,leuschner2018fourier,ladecky2023optimal}. 
We consider a cell $Y$ discretized by $N_x\times N_y\times N_z$ voxels. The vector ${\bm{\gamma}}$ comprises the integer frequencies of the discrete Fourier transform and runs through the index set
\begin{equation}\label{eq:frequencySet}
    \mathcal{Z}_N = \{0,\dots,N_x-1\}\times\{0,\dots,N_y-1\}\times\{0,\dots,\lfloor N_z/2\rfloor\}.
\end{equation}
The third direction is halved, since the Fourier coefficients of a real-valued field are Hermitian symmetric, and a real-to-complex FFT stores only the non-redundant half. The matrix, which represents the action of the SBFEM Green's operator in Fourier space, is computed using the formula
\begin{equation}\label{eq:GreenHermitian}
    {\widehat{\mathbf{K}_0}}^{-1}({\bm{\gamma}}) = \begin{cases}
    \left(\left({\mathbf{\Theta}}({\bm{\gamma}})\right)^{H} {\mathbf{K}_0^e}\,{\mathbf{\Theta}}({\bm{\gamma}})\right)^{-1}, & {\bm{\gamma}}\in\mathcal{Z}_N\setminus\{{\mathbf{0}}\},\\
    {\mathbf{0}}, & {\bm{\gamma}}={\mathbf{0}},
    \end{cases}
\end{equation}
where $H$ denotes the Hermitian conjugate of a complex-valued matrix, which consists of complex conjugation and transposition. The matrices~\eqref{eq:GreenHermitian} depend neither on the material nor on the iterate. Thus, following Ladecký et al.~\cite{ladecky2023optimal}, we precompute them once for all frequencies ${\bm{\gamma}}\in\mathcal{Z}_N$ and cache them throughout the iterations.
The complex-valued matrix ${\mathbf{\Theta}}$ of dimension $24\times3$, appearing in Eq.~\eqref{eq:GreenHermitian}, takes the form
\begin{equation}
    {\mathbf{\Theta}}({\bm{\gamma}}) = {\bm{\varphi}}({\bm{\gamma}})\otimes{\mathbf{I}_3}
\end{equation}
and encodes the relative position of the hexahedral element nodes within a voxel based on the Kronecker product $\otimes$, the $3\times3$ identity matrix ${\mathbf{I}_3}$ and the complex-valued vector ${\bm{\varphi}}$ with the eight entries
\begin{equation}\label{eq:phaseVector}
   \left({\bm{\varphi}}\right)_{j}({\bm{\gamma}}) = \exp\left(\frac{2\pi i}{h}\,(\widehat{\fx}_j-\widehat{\fx}_1)\cdot{\bm{\vartheta}}({\bm{\gamma}})\right)\quad\text{for}\quad j=1,\ldots,8
\end{equation}
with the complex unit $i$, the corner nodes $\widehat{\fx}_j$ of the reference voxel~\eqref{eq:refVoxel}, numbered as in Fig.~\ref{fig:elementUwithEps}, and the normalized frequency vector
\begin{equation}\label{eq:normalizedFrequency}
   {\bm{\vartheta}}({\bm{\gamma}}) = \left(\gamma_x/N_x,\ \gamma_y/N_y,\ \gamma_z/N_z\right).
\end{equation}
We note that the inverse in Eq.~\eqref{eq:GreenHermitian} is a standard inverse, since for non-zero frequency \mbox{${\bm{\gamma}} \neq {\mathbf{0}}$}, the range of the matrix ${\mathbf{\Theta}}$ contains no rigid body modes and thus lies outside the kernel of the homogeneous element stiffness matrix~\eqref{eq:KIElement}. At zero frequency, the Fourier coefficient of a nodal field coincides, up to the normalization of the FFT, with its cell average. Setting the SBFEM Green's operator~\eqref{eq:GreenHermitian} to zero at ${\bm{\gamma}}={\mathbf{0}}$ therefore removes the average of the displacement update and enforces the mean-free condition on the displacement fluctuation, whereas the effective strain $\bfeps$ enters via the affine part of the displacement field. On the space of all nodal vectors, the matrices ${\mathbf{K}}$ and ${\mathbf{K}_0}$ share a three-dimensional kernel comprising the rigid translations. Thus, the operator~\eqref{eq:GreenHermitian} realizes the (Moore-Penrose) pseudo-inverse of the matrix ${\mathbf{K}_0}$, which coincides with its inverse on the nodal vectors with vanishing mean.

The FFT-based solution scheme is aborted once the stopping criterion~\cite{schneider2021review}
\begin{equation}
    \mathrm{res}_k \stackrel{!}{\leq} \mathrm{tol} \, \norm{\left\langle \fsigma({\mathbf{u}}^k)\right\rangle_Y}
    \label{eq:stoppingCriterion}
\end{equation}
is satisfied for a prescribed tolerance $\mathrm{tol}$, where the average stress at iteration $k$ is computed using Eq.~\eqref{eq:normSigma}.
The residual per iteration $k$ is computed by the relation~\cite{schneider2022voxel}
\begin{equation}
    \text{res}_k = \sqrt{\T{{\mathbf{r}}({\mathbf{u}}^k)} \, {\mathbf{K}_0}^{-1} \, {\mathbf{r}}({\mathbf{u}}^k)/|Y|},
    \label{eq:residual}
\end{equation}
with the cell volume $|Y|$ defined in Section~\ref{sec:PeriodicHomogenization}, where the SBFEM Green's operator is applied in Fourier space using the representation~\eqref{eq:GreenHermitian} and a Fourier transformed version of the global nodal force vector~\eqref{eq:nodalRes}.

\begin{algorithm}
\caption{Displacement-based basic scheme}\label{alg: basicSchemeSBFEM}
\begin{algorithmic}[1]
\State Determine the step size \,$s$ by Eq.~\eqref{eq:stepsize}.
\State $\left({\mathbf{u}},{\mathbf{g}}, k\right) \gets \left({\mathbf{0}}, {\mathbf{0}}, 0\right)$  \Comment{Initialization.}
\State$\mathrm{res} \gets \mathrm{tol} + 1$
\While{$k<\mathrm{maxit}$ and $\mathrm{res}>\mathrm{tol} \norm{\left\langle \fsigma\right\rangle_Y}$ }\Comment{Average stress via Eq.~\eqref{eq:normSigma}.}
\State $k \gets k+1$
\State {${\widehat{\mathbf{g}}}\gets \fourier{{\mathbf{r}}({\mathbf{u}})}$}\Comment{Using Eq.~\eqref{eq:nodalRes}.}
\State {${\mathbf{g}} \gets  \invfourier{\widehat{ {\mathbf{K}_0}}^{-1} \,{\widehat{\mathbf{g}}}}$} \Comment{Using expression~\eqref{eq:GreenHermitian}.}
\State {$ \mathrm{res} \gets  \sqrt{\T{{\mathbf{r}}({\mathbf{u}})}\,{\mathbf{g}}/|Y|}$} 
\State ${\mathbf{u}} \gets {\mathbf{u}} - s\,{\mathbf{g}}$
\EndWhile
\end{algorithmic}
\end{algorithm}

Following the established displacement-based implementations~\cite{leuschner2018fourier,schneider2022voxel,ladecky2023optimal}, the displacement-based implementation of the SBFEM basic scheme, discussed in this section, operates without assembling the global stiffness matrix. Instead, all matrix-vector products are computed locally on the hexahedral elements or nodes and are thus readily parallelized. The resulting memory consumption scales linearly with the number of voxels, since we solely store one element stiffness matrix per material, the precomputed Green's operator~\eqref{eq:GreenHermitian} and two global nodal displacement vectors.
We summarize the implementation by discussing the pseudocode shown in Alg.~\ref{alg: basicSchemeSBFEM}. 
After initialization in lines 1-3, we enter a while loop, which is only terminated once the stopping criterion~\eqref{eq:stoppingCriterion} is reached. For each iteration $k$, we compute the nodal residual vector and transform it to Fourier space in line 6. In line 7, the SBFEM Green's operator is applied in Fourier space and the result is transformed back to real space. Line 8 evaluates the residual~\eqref{eq:residual}, and the displacement fluctuation is updated in line 9. All vectors and matrices that are represented in Fourier space are marked by a hat superscript in Alg.~\ref{alg: basicSchemeSBFEM}.

The discussed SBFEM basic scheme forms the basis of fast gradient methods, such as the Barzilai-Borwein scheme~\cite{schneider2019barzilai}, the linear conjugate gradient method~\cite{zeman2010accelerating} or the nonlinear conjugate gradient method~\cite{schneider2020dynamical}. For a discussion on the required modifications towards these schemes, we refer to Schneider~\cite[§3]{schneider2022voxel}.

\section{Computational investigations}\label{sec:Experiments}
\subsection{Setup}\label{sec:setup}

\begin{figure}[htb]
  \centering
  \begin{subfigure}[t]{0.32\textwidth}
      \includegraphics[width=\textwidth]{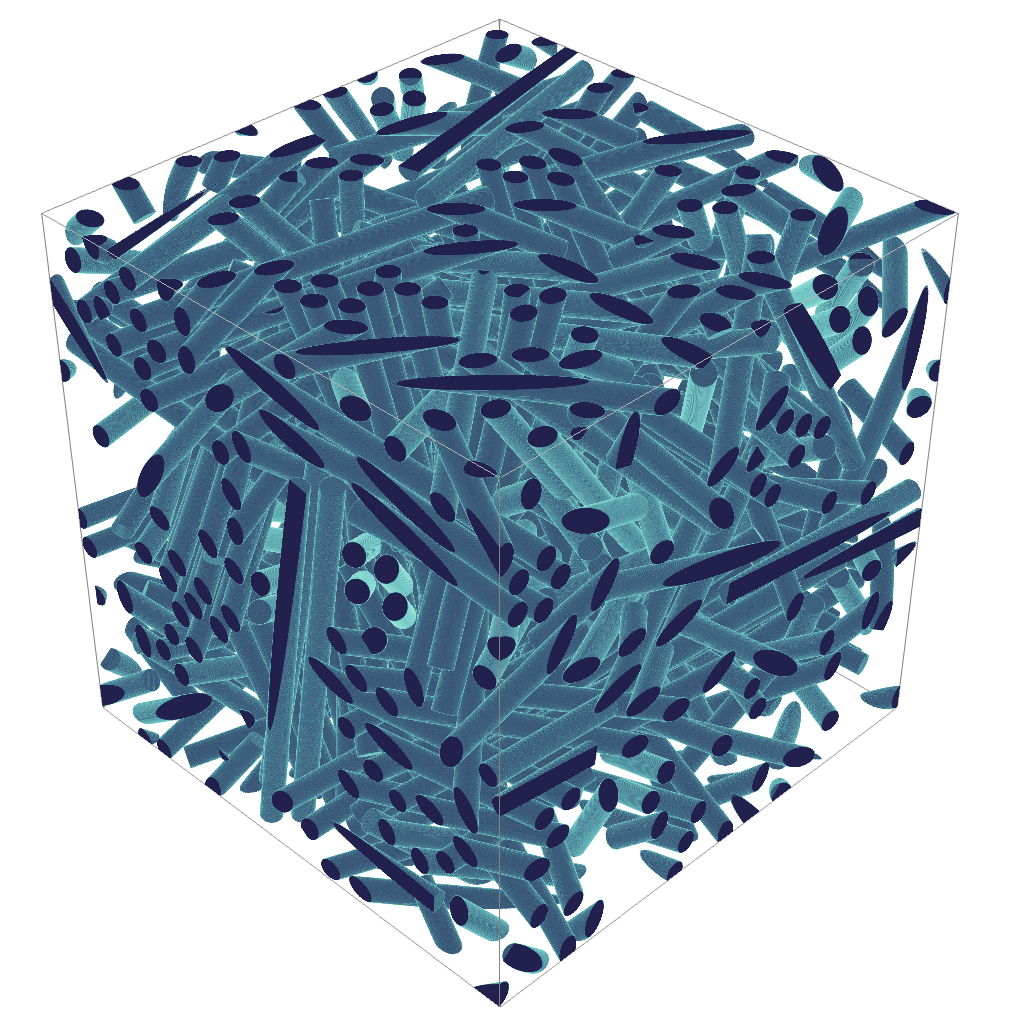}
      \caption{Short fibre reinforced composite}\label{fig:microstructuresFiber}
  \end{subfigure}  
  \begin{subfigure}[t]{0.32\textwidth}
      \includegraphics[width=\textwidth]{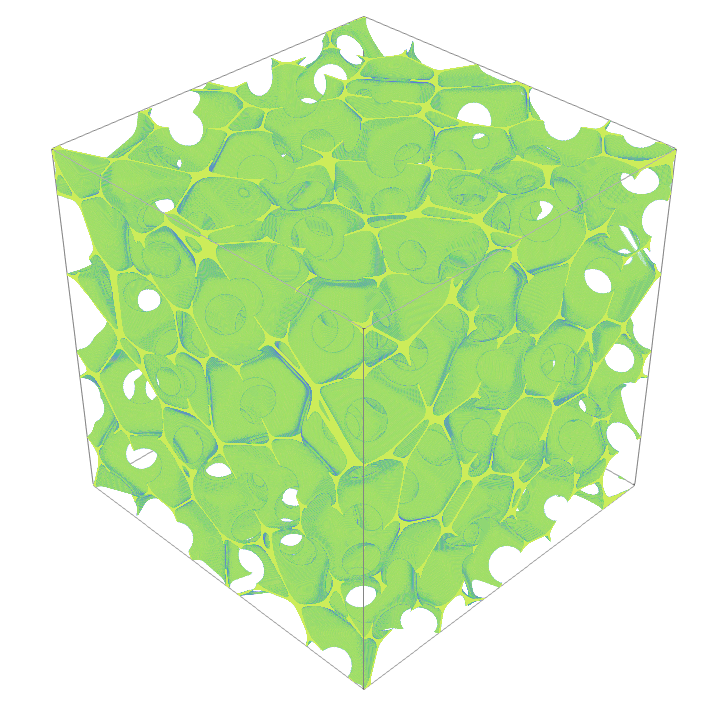}
      \caption{Polyurethane foam}\label{fig:microstructuresFoam}
  \end{subfigure}  
  \begin{subfigure}[t]{0.32\textwidth}
      \includegraphics[width=\textwidth]{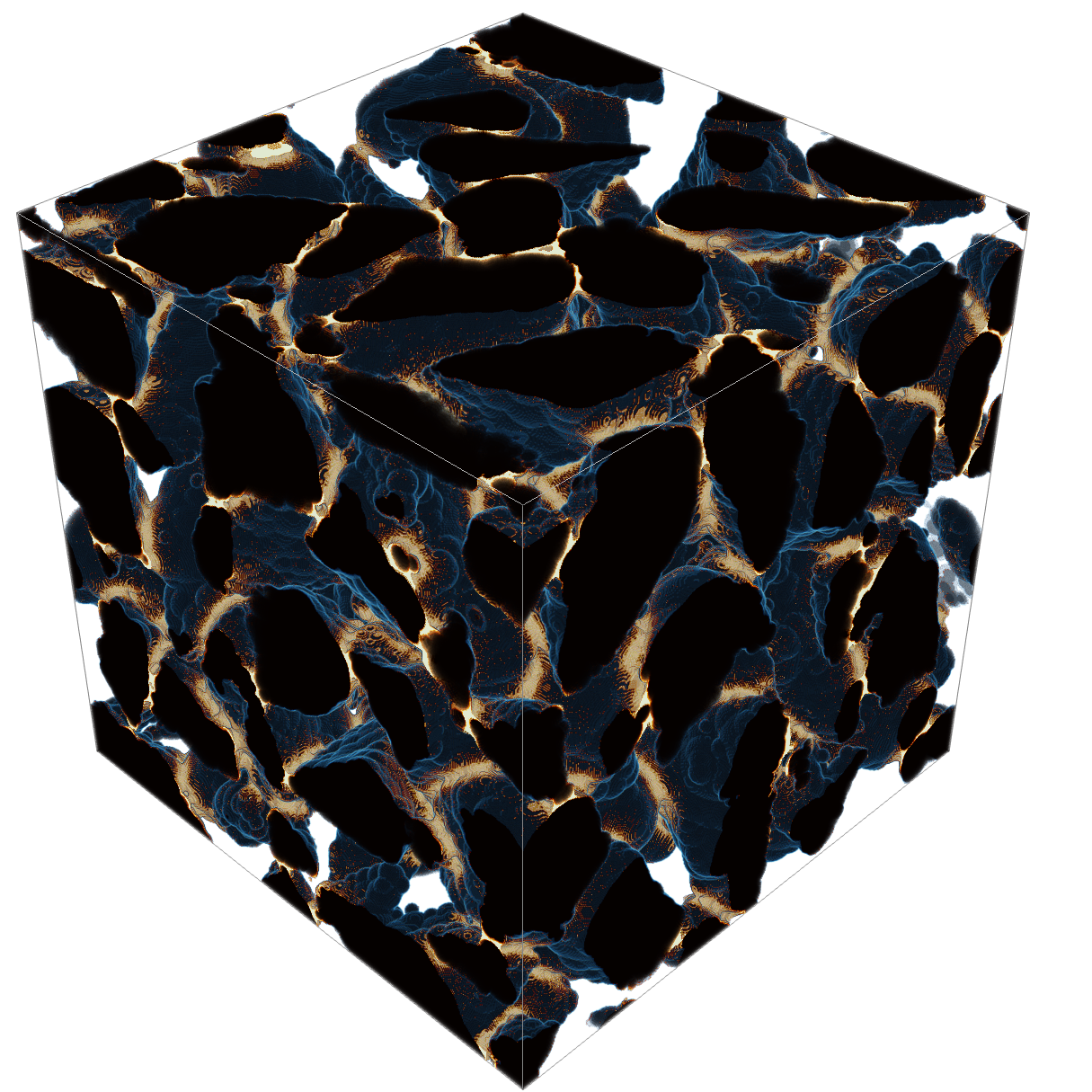}
      \caption{Sand-based binder jet 3D printed microstructure}\label{fig:microstructuresSand}
  \end{subfigure}
\caption{Geometries of the investigated microstructures.}\label{fig:microstructures}
\end{figure}

We assess the accuracy and the numerical efficiency of the SBFEM-FFT method on linear elastic homogenization problems for three microstructures, shown in Fig.~\ref{fig:microstructures}: a short-fibre reinforced composite with finite material contrast, a polyurethane foam and a sand-based binder jet 3D printed microstructure, the latter two with pores. The linear elastic material parameters are given in Tab.~\ref{tab:material}.

\begin{table}[htb]
        \centering
        \begin{tabular}{l|lllll}
          & Polypropylene&E-glass&Polyurethane&Sand&Binder\\
          \hline
        Young's modulus&$1.25\,\text{GPa}$~\cite{dow2003c711}&$72 \,\text{GPa}$~\cite{glass2013tufrov}&$2.5 \,\text{GPa}$~\cite{goods1999mechanical}&$66.9\,\text{GPa}$~\cite{daphalapurkar2011determination}&$43\,\text{GPa}$~\cite{lechner2020novel}\\
        Poisson's ratio&$0.35$~\cite{dow2003c711}&$0.22$~\cite{glass2013tufrov}&$0.35$~\cite{goods1999mechanical}&$0.25$~\cite{wichtmann2010influence}&$0.17$~\cite{sanditov2009poisson}\\
        \end{tabular}
        \caption{Material parameters.}
\label{tab:material}
\end{table}

We compare the results of the SBFEM discretization with the SBFEM-FFT preconditioner derived in this work with those of the following other discretizations with corresponding FFT-based preconditioner:
\begin{enumerate}
    \item trilinear hexahedral finite elements (HEX8)~\cite{schneider2017fft,leuschner2018fourier,ladecky2023optimal}
    \item rotated staggered grid~\cite{willot2015fourier} discretization, which may be interpreted~\cite{schneider2017fft} as trilinear hexahedral finite elements with reduced integration (HEX8R)
    \item tetrahedron-based finite differences (TET)~\cite{finel2025tetrahedron}.
\end{enumerate}
We apply a uniaxial strain loading with the effective strain $\bar{\eps}_{11} = 0.1\%$ and vanishing remaining components of the effective stress, and investigate the accuracy of the effective stress response as an indicator of how accurately the effective stiffness of the microstructure cell is approximated by the different discretizations. Moreover, we analyse the numerical efficiency of the different discretizations by examining the required iteration count until convergence of the solver and runtime of the simulation.
All simulations were performed using our in-house displacement-based FFT software, which is implemented in Python with Cython extensions and parallelized using OpenMP. Each study is run on a workstation with 1.15 TB RAM and 96 CPU cores at $2.4\,$GHz using 16 threads per simulation.

\subsection{Short fibre reinforced composite}
\begin{figure}
  \centering
    \begin{subfigure}[t]{0.5\textwidth}
        \fbox{\includegraphics[width=\dimexpr\textwidth-2\fboxsep-2\fboxrule\relax]{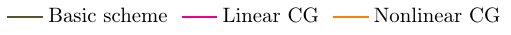}}
    \end{subfigure}\\
    \begin{subfigure}[t]{0.32\textwidth}
        \centering 
        \includegraphics[width=\textwidth]{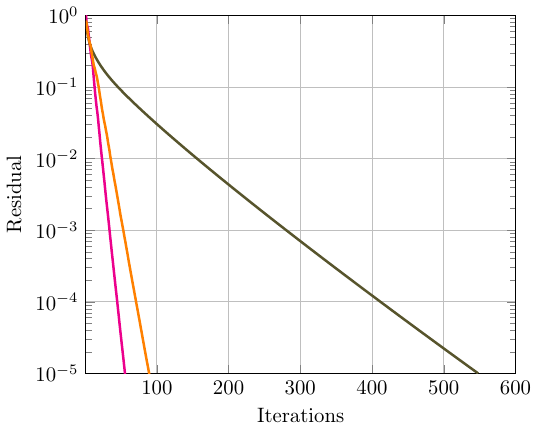}
        \caption{Short fibre reinforced composite}\label{fig:solverFiber}
    \end{subfigure}
    \begin{subfigure}[t]{0.32\textwidth}
        \centering 
        \includegraphics[width=\textwidth]{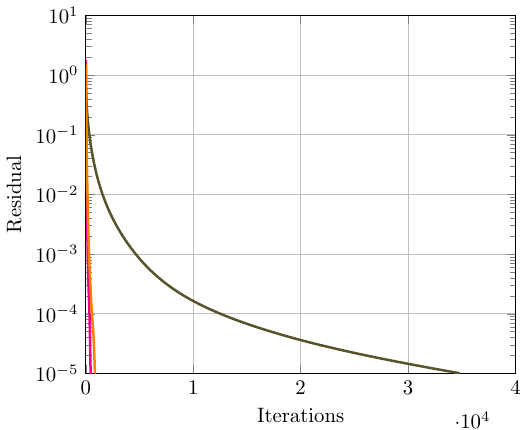}
        \caption{Polyurethane foam}\label{fig:solverFoam}
    \end{subfigure}
    \begin{subfigure}[t]{0.32\textwidth}
        \centering 
        \includegraphics[width=\textwidth]{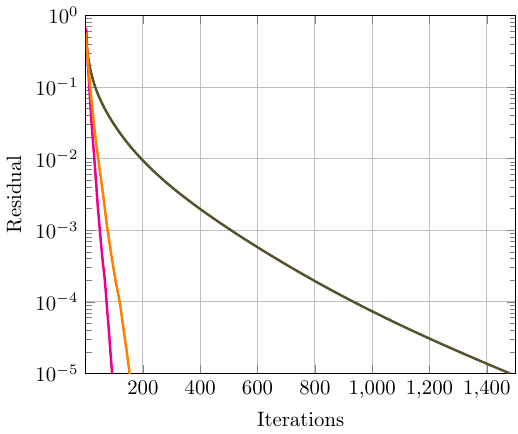}
        \caption{Sand-based binder jet 3D printed microstructure}\label{fig:solverSand}
    \end{subfigure}
\caption{Solver convergence of SBFEM at a resolution of $128^3$ voxels.}\label{fig:solver}
\end{figure}

First, we investigate a short fibre reinforced composite, consisting of a polypropylene matrix and E-glass fibres with the linear elastic properties listed in Tab.~\ref{tab:material}.
The microstructure has an edge length of $256\mu$m along all edges and a fibre volume fraction of 20\%, with a fibre length of $200\mu$m and a diameter of $10\mu$m. We generated the microstructure, shown in Fig.~\ref{fig:microstructuresFiber}, using the sequential addition and migration method~\cite{schneider2017sequential}. A uniaxial strain loading with $\bar{\eps}_{11} = 0.1\%$ is prescribed.

In Fig.~\ref{fig:solverFiber}, the solver convergence of the SBFEM discretization with derived preconditioner is analysed for the basic scheme~\cite{moulinec1994fast,moulinec1998numerical}, the linear CG scheme~\cite{zeman2010accelerating}, and the nonlinear CG scheme~\cite{schneider2020dynamical}. We terminate the solver as soon as the residual is smaller than the prescribed tolerance of $10^{-5}$. For the short fibre reinforced composite study, we observe that the linear CG scheme requires the fewest iterations, namely $57$, to reduce the residual~\eqref{eq:residual} below the tolerance. Nonlinear CG takes $91$ iterations, whereas the basic scheme needs $549$. We therefore use linear CG for all subsequent studies in this section.

\begin{figure}[htb]
  \centering
	\fbox{\includegraphics[width=0.6\textwidth]{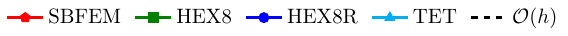}}\\
   \begin{subfigure}[t]{0.32\textwidth}
        \centering 
        \includegraphics[width=\textwidth]{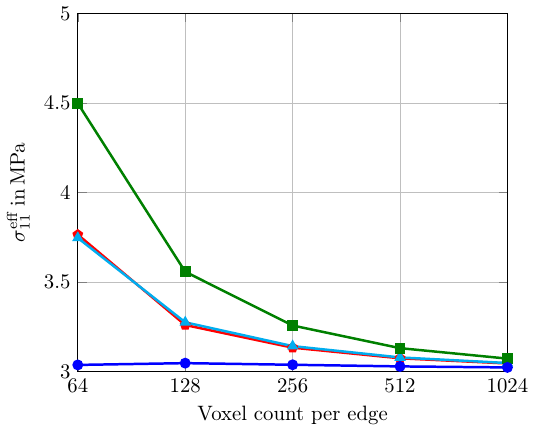}
        \subcaption{Effective stress} \label{fig:EffStressShortFiber}
    \end{subfigure}
       \begin{subfigure}[t]{0.32\textwidth}
        \centering 
        \includegraphics[width=\textwidth]{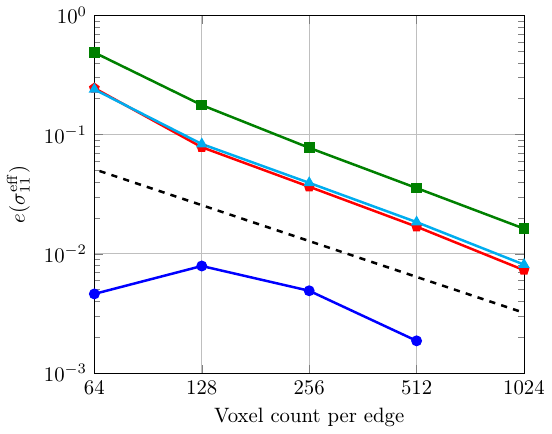}
        \subcaption{Error convergence} \label{fig:ErrEffStressShortFiber}
    \end{subfigure}
    \begin{subfigure}[t]{0.30\textwidth}
        \centering 
        \includegraphics[width=\textwidth]{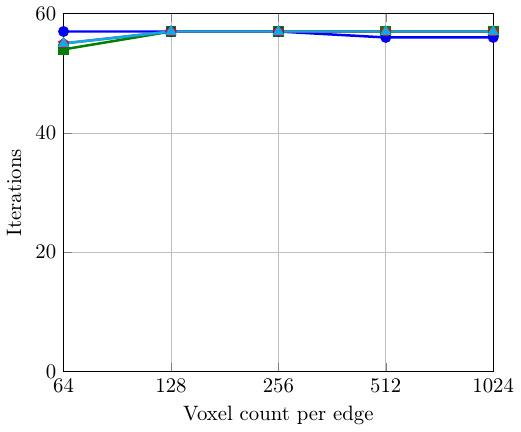}
        \subcaption{Iteration count}\label{fig:iterationsFiber}
    \end{subfigure}
\caption{Resolution study for the short fibre reinforced composite using the linear CG scheme and a uniaxial strain loading with $\bar{\eps}_{11} = 0.1\%$.}
\end{figure}

To assess the accuracy, we examine the effective stress $\sigma_{11}$ versus the voxel count per edge in Fig.~\ref{fig:EffStressShortFiber}. As the resolution increases, i.e., as the voxel count per edge increases, the effective stresses of all discretizations appear to converge to the same value. However, as previously noted~\cite{schneider2022voxel, gehrig2025element}, HEX8 leads to an overly stiff material response at low resolutions. We observe that, at low resolutions, HEX8R approximates the effective stress $\sigma_{11}$ most accurately. SBFEM yields results of the effective stress that lie between those of HEX8 and HEX8R for all resolutions considered. In particular, the SBFEM stresses lie below the HEX8 stresses at every resolution, as predicted by the energy ordering~\eqref{eq:energyOrdering}. The ordering of the effective stiffnesses reverses for the effective compliances and thus carries over to the effective stress $\sigma_{11}$ under the uniaxial strain loading at hand. The values of TET are close to those of SBFEM for all considered discretization levels.

Fig.~\ref{fig:ErrEffStressShortFiber} shows the relative error in the effective stress with respect to the HEX8R stress at voxel count $1024^3$, which serves as the reference. For TET, HEX8 and SBFEM, the error in the effective stress decreases linearly in the grid spacing $h$. 
For moderate resolutions, HEX8R also features a linear error convergence rate, however, we observe an outlier at voxel count $64^3$.
We compute the effective stresses of the SBFEM via the exact element averages~\eqref{eq:stressAverage}~\cite{wang2017efficient}.

We study the numerical efficiency of the different discretizations. As shown in Fig.~\ref{fig:iterationsFiber}, the linear CG solver requires between $54$ and $57$ iterations for all discretizations and resolutions. Tab.~\ref{tab:runtimeFiber} lists the respective runtimes. 
For the study at hand, the runtimes of the different discretizations are rather similar at a specific resolution. At a voxel count of $128^3$ voxels, the FFT-based framework solves the problem with linear CG in less than 25 seconds for all discretizations, using the hardware specifications listed in Section~\ref{sec:setup}. A resolution of $1024^3$ voxels, which corresponds to over three billion degrees of freedom requires around two hours and a peak memory consumption of $139-178$ GB of RAM to compute for every discretization.

\begin{table}[htb]
        \centering
        \begin{tabular}{l|llll}
         Voxel count&SBFEM&HEX8&HEX8R&TET\\
          \hline
        $128^3$&$20.5\,$s&$22.8\,$s&$22.1\,$s&$22.1\,$s\\
        $256^3$&$2.4\,$min&$2.7\,$min&$2.6\,$min&$2.7\,$min\\
        $512^3$&$16.6\,$min&$17.8\,$min&$17.9\,$min&$18.0\,$min\\
        $1024^3$&$1.8\,$h&$1.9\,$h&$1.9\,$h&$1.9\,$h\\
        \end{tabular}
        \caption{Runtimes for the short fibre reinforced composite using the linear CG scheme with 16 threads and the hardware specifications listed in Section~\ref{sec:setup}.}\label{tab:runtimeFiber}
\end{table}

To conclude, for the short fibre reinforced composite, as an example of a solid microstructure with a material contrast~\eqref{eq:materialContrast} of around 139, HEX8R provides the best ratio between accuracy and numerical efficiency. SBFEM provides a comparable accuracy and numerical effort to TET. HEX8 strongly overpredicts the effective stress at low resolutions.

\subsection{Polyurethane foam}\label{sec:polyurethaneFoam}
This section is dedicated to a polyurethane foam as an example of a highly porous microstructure with complex pore space.
Polyurethane foams combine a low density and a low thermal conductivity with favourable mechanical properties, which makes them suitable as thermal and acoustic insulators as well as structural and comfort materials~\cite{gama2018polyurethane,ates2022polyurethane}. Fig.~\ref{fig:microstructuresFoam} shows a polyurethane foam microstructure with a connected pore space. Some of its cell windows are closed by thin membranes. Such partially reticulated foams commonly occur in practice~\cite{gholami2017effect}. The microstructure has a solid volume fraction of 8\% and an edge length of $1.28\,$mm. It represents porous microstructures with thin cell walls. We use the material parameters of polyurethane listed in Table~\ref{tab:material} and prescribe a uniaxial strain loading with $\bar{\eps}_{11} = 0.1\%$.

In a preliminary study, we analyse the solver convergence of the SBFEM discretization with derived preconditioner in Fig.~\ref{fig:solverFoam}. For the stopping criterion~\eqref{eq:stoppingCriterion}, we prescribe a tolerance of $10^{-5}$. We observe that the linear CG scheme~\cite{zeman2010accelerating} requires the fewest iterations, namely $451$, to reduce the residual~\eqref{eq:residual} below the tolerance. Nonlinear CG~\cite{schneider2020dynamical} takes $857$ iterations, whereas the basic scheme~\cite{moulinec1994fast,moulinec1998numerical} needs more than $30\,000$. Based on these results, we use the linear CG scheme for this section.
For all three schemes, the residual eventually decreases by a constant factor per iteration, i.e., the schemes converge linearly. For the basic scheme and linear CG, this behaviour agrees with the theory for porous microstructures~\cite{schneider2020lippmann}.

\begin{figure}[htb]
  \centering
	\fbox{\includegraphics[width=0.4\textwidth]{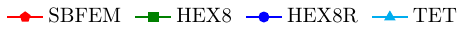}}\\
   \begin{subfigure}[c]{0.45\textwidth}
        \centering 
        \includegraphics[width=\textwidth]{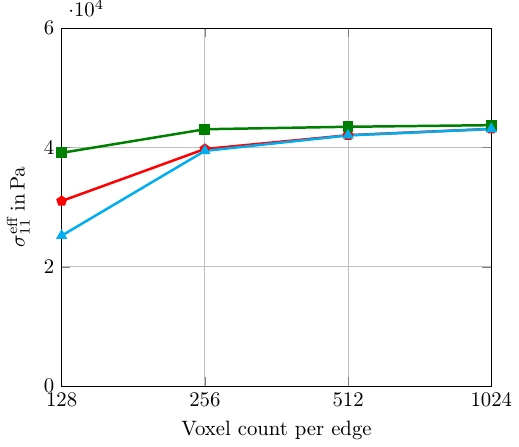}
        \subcaption{Effective stress} \label{fig:EffStressFoam}
    \end{subfigure}
    \begin{subfigure}[c]{0.45\textwidth}
        \centering 
        \includegraphics[width=\textwidth]{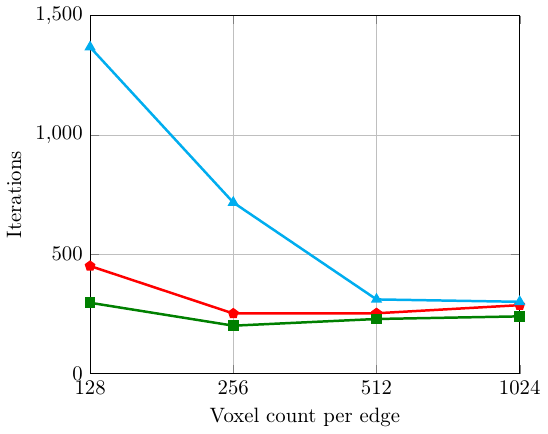}
        \subcaption{Iteration count}\label{fig:iterationsFoam}
    \end{subfigure}
\caption{Resolution study for the polyurethane foam using the linear CG scheme and a uniaxial strain loading with $\bar{\eps}_{11} = 0.1\%$.}
\end{figure}

\begin{table}[htb]
        \centering
        \begin{tabular}{l|llll}
         Voxel count&SBFEM&HEX8&HEX8R&TET\\
          \hline
        $128^3$&$2.8\,$min&$2.3\,$min&-&$9.6\,$min\\
        $256^3$&$11.1\,$min&$9.8\,$min&-&$34.5\,$min\\
        $512^3$&$1.4\,$h&$1.3\,$h&-&$1.8\,$h\\
        $1024^3$&$8.7\,$h&$7.7\,$h&-&$9.7\,$h\\
        \end{tabular}
        \caption{Runtimes for the polyurethane foam using the linear CG scheme with 16 threads and the hardware specifications listed in Section~\ref{sec:setup}.}\label{tab:runtimeFoam}
\end{table}
Fig.~\ref{fig:EffStressFoam} shows the effective stress $\sigma_{11}$ versus the voxel count per edge. We observe that HEX8, TET, and SBFEM converge to approximately identical stress values at 1024 voxels per edge. At low resolution, i.e., at low voxel count per edge, we observe that HEX8 approximates the effective stress most accurately, followed by SBFEM and TET. This observation does not contradict the energy ordering~\eqref{eq:energyOrdering}, which compares the discretizations on the same voxelized microstructure. Indeed, the SBFEM stresses lie below the HEX8 stresses at every resolution. At low resolutions, however, the voxelization alters the thin cell walls, and the comparison with the stress at $1024$ voxels per edge includes this geometric error. Thus, the overly stiff response of HEX8 partially compensates the geometric error, which explains its apparent accuracy. HEX8R did not converge with less than $50\,000$ iterations for all solvers and is therefore not shown in Fig.~\ref{fig:EffStressFoam}. The numerical instability of the HEX8R element for voids with complex pore space is well-documented in the literature~\cite{schneider2020lippmann} and is attributed to hourglass instabilities. To address this issue, a suitable remedy is to adapt the HEX8R discretization using hourglass control techniques~\cite{schneider2022voxel}.

In Fig.~\ref{fig:iterationsFoam}, we show the iteration count until convergence of linear CG versus the resolution.
Notably, HEX8 finishes with the least number of iterations across all resolutions for the polyurethane foam. In contrast, SBFEM requires a slightly higher iteration count, whereas TET shows the largest iteration count among the convergent discretizations. The iteration count of TET is particularly high at low resolutions.

The runtimes shown in Tab.~\ref{tab:runtimeFoam} generally follow the trends observed in the iteration count. We highlight that, for a polyurethane foam microstructure, which is discretized by $256^3$ voxels, the FFT-based framework solves the problem with SBFEM in around 11 minutes.

All in all, HEX8 shows the highest accuracy for the lowest numerical effort for the polyurethane foam microstructure. HEX8R did not converge for this microstructure due to hourglass instabilities occurring for voids with complex pore space. SBFEM has a similar approximation accuracy to TET, but requires lower numerical effort for the polyurethane foam.

\subsection{Sand-based binder jet 3D printed microstructure}

\begin{figure}[htb]
  \centering
	\fbox{\includegraphics[width=0.4\textwidth]{legendStudies.pdf}}\\
   \begin{subfigure}[c]{0.45\textwidth}
        \centering 
        \includegraphics[width=\textwidth]{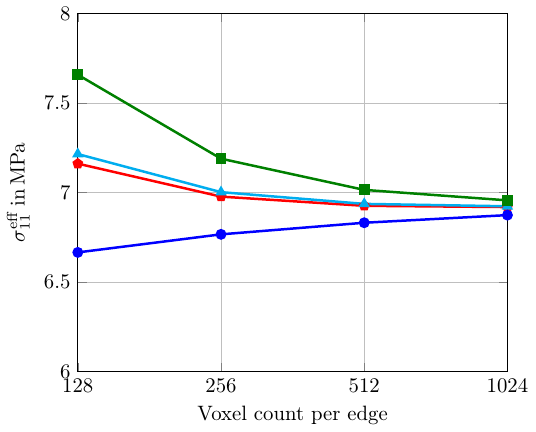}
        \subcaption{Effective stress} \label{fig:EffStressSand}
    \end{subfigure}
    \begin{subfigure}[c]{0.45\textwidth}
        \centering 
        \includegraphics[width=\textwidth]{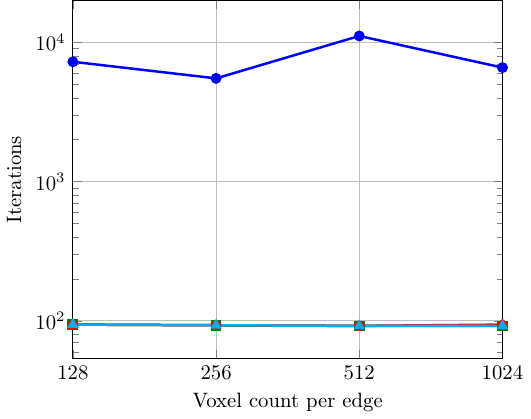}
        \subcaption{Iteration count}\label{fig:IterationsSand}
    \end{subfigure}
\caption{Resolution study for the sand-based binder jet 3D printed microstructure.}
\end{figure}

Finally, we consider a sand-based binder jet 3D printed microstructure as an example of a three-phase material, including pores. The sand-based binder jet 3D printed microstructure is shown in Fig.~\ref{fig:microstructuresSand}. It has an edge length of $128\mu\text{m}$ along all edges and comprises 64 grains with average grain diameter $32\mu$m. The microstructure features 51 vol-\% of sand grains and 2 vol-\% of binder, and was generated using a directional contraction method~\cite{donval2025directional}. We use the material properties of the sand grains and the binder that are listed in Tab.~\ref{tab:material}.

In a preliminary study, we analyse the solver convergence of the SBFEM discretization with a prescribed residual tolerance of $10^{-5}$. In Fig.~\ref{fig:solverSand}, we observe that the linear CG scheme~\cite{zeman2010accelerating} reaches the tolerance after $94$ iterations, whereas the nonlinear CG scheme~\cite{schneider2020dynamical} takes $155$ and the basic scheme $1482$ iterations. We therefore use the linear CG scheme for all subsequent computations in this section.

Fig.~\ref{fig:EffStressSand} shows the effective stress $\sigma_{11}$ versus the voxel count per edge. At the highest resolution of $1024$ voxels per edge, all discretizations lead to approximately the same stress value. At low resolution, HEX8 overpredicts the effective stress, although less than for the short fibre reinforced composite in Fig.~\ref{fig:EffStressShortFiber}. In accordance with the energy ordering~\eqref{eq:energyOrdering}, the SBFEM stresses lie below the HEX8 stresses at every resolution. SBFEM and TET show again rather similar values at all resolutions. Together with the HEX8R discretization, they appear to lead to the most accurate stress values at low resolution.

The iteration count until convergence of linear CG is shown in Fig.~\ref{fig:IterationsSand}. HEX8, TET and SBFEM require between $92$ and $94$ iterations for all resolutions. In contrast, the iteration count of HEX8R is between $5\,500$ and $11\,100$ iterations. The high iteration count of HEX8R indicates a high condition number of the preconditioned system, that is a consequence of the void phase of the sand-based binder jet 3D printed microstructure and is caused by hourglass instabilities, as discussed in Sec.~\ref{sec:polyurethaneFoam}.

\begin{table}
        \centering
        \begin{tabular}{l|llll}
         Voxel count&SBFEM&HEX8&HEX8R&TET\\
          \hline
        $128^3$&$38.2\,$s&$46.4\,$s&$51.6\,$\textbf{min}&$49.6\,$s\\
        $256^3$&$4.0\,$min&$4.7\,$min&$4.1\,$\textbf{h}&$5.0\,$min\\
        $512^3$&$27.1\,$min&$29.0\,$min&$2.2\,$\textbf{d}&$29.4\,$min\\
        $1024^3$&$3.4\,$h&$3.6\,$h&$9.2\,$\textbf{d}&$3.6\,$h\\
        \end{tabular}
        \caption{Runtimes for the sand-based binder jet 3D printed microstructure using the linear CG scheme with 16 threads and the hardware specifications listed in Section~\ref{sec:setup}.}\label{tab:runtimeSand}
\end{table}

The runtimes until convergence of linear CG are shown in Tab.~\ref{tab:runtimeSand}. SBFEM, TET and HEX8 show runtimes of similar order of magnitude at all resolutions: For sand-based binder jet 3D printed microstructure that is resolved by $128^3$ voxels, SBFEM, TET and HEX8 require less than one minute, at voxel count $1024^3$ they require less than four hours. The runtimes of HEX8R are exceptionally high, due to the conditioning problems that result from the pore space of the sand-based binder jet 3D printed microstructure.

All in all, for the implementation at hand, SBFEM and TET show the best balance between accuracy and numerical effort for the sand-based binder jet 3D printed microstructure. HEX8 overpredicts the effective stress at low resolutions, whereas HEX8R experiences numerical instabilities due to hourglass modes.

\section{Conclusions}
\label{sec:conclusion}

In this work, we introduced a voxel-based SBFEM discretization for FFT-based computational homogenization of linear elastic materials. The SBFEM constructs the displacement field within each voxel from a discretization of its boundary, which is scaled towards the centre of the voxel. The radial variation follows from the analytical solution of the scaled boundary finite element equation, such that the shape functions depend on the material of the voxel. Thus, the SBFEM is material-adaptive: Within each voxel, the radial displacement minimizes the elastic energy among all radial displacements with the same nodal displacements. This principle resembles the interpretation of composite voxels as an assumed enhanced strain method~\cite{lendvai2024assumed}, where the strain field likewise adapts to the material within the voxel.

For the FFT-based homogenization community, the SBFEM provides a voxel discretization that combines the strengths of the established hexahedral elements. Owing to its material-dependent shape functions, the SBFEM is softer than fully integrated hexahedral elements, and the energy ordering~\eqref{eq:energyOrdering} guarantees this property on every voxel grid. At the same time, the SBFEM is a conforming Galerkin discretization and does not suffer from the hourglass instabilities that render reduced-integration hexahedral elements, i.e., the rotated staggered grid, unreliable for porous microstructures with complex pore space. Across the investigated microstructures, the SBFEM provided a robust compromise between the two hexahedral discretizations, and for the polyurethane foam, it achieved a better balance between accuracy and efficiency than the tetrahedron-based discretization~\cite{finel2025tetrahedron}. The material-dependent shape functions required an argument of their own for the preconditioner. As the semi-analytical space does not depend on the material, the SBFEM-FFT preconditioner admits a local representation in Fourier space and a bound on the condition number that depends only on the material contrast.

For the SBFEM community, the coupling to FFT-based solvers makes large voxel images accessible without an octree decomposition of the image and without relying on black-box solvers. The matrix-free implementation stores one element stiffness matrix per material and the precomputed Green's operator. Even at a resolution of $1024^3$ voxels, corresponding to roughly three billion degrees of freedom, the SBFEM-FFT simulations took between $1.8$ and $8.7$ hours using $16$ threads and less than $180\,$GB of RAM on a single CPU workstation. With the exact element averages~\eqref{eq:stressAverage}, the error in the effective stress decreases linearly in the grid spacing.

The SBFEM-FFT framework may serve as a baseline for several extensions. For a start, an SBFEM counterpart of composite voxels~\cite{gelebart2015filtering,kabel2015use,lendvai2024assumed} would be desirable to account for material interfaces within a voxel. Secondly, the SBFEM provides an upper bound on the effective stiffness of the voxelized microstructure that is sharper than the bound of fully integrated hexahedral elements, as stated by the energy ordering~\eqref{eq:energyOrdering}. Combining it with a lower bound from a complementary, stress-based discretization~\cite{vondvrejc2015guaranteed} would yield guaranteed two-sided bounds on the effective properties. Thirdly, adaptive refinement~\cite{kabel2024adaptive} and the resolution of cracks via enriched phase-field methods~\cite{Loehnert2023XPFM} appear feasible on top of the SBFEM-FFT framework. For inelastic materials, the condensed space may be constructed once with the elastic stiffness and kept fixed during the computation, which retains material-adapted shape functions without repeated condensation, possibly combined with element-based strategies for reducing the number of material law evaluations~\cite{gehrig2025element}. Last but not least, the preconditioner argument carries over to scalar problems such as heat conduction, and the regular, matrix-free structure of the implementation is well suited for graphics processing units.

\section*{CRediT authorship contribution statement}
\textbf{Flavia Gehrig:} Conceptualization, Methodology, Formal analysis, Software, Validation, Investigation, Data curation, Visualization, Writing -- original draft, Writing -- review \& editing. \textbf{Carolin Birk:}  Conceptualization, Methodology, Formal analysis, Visualization, Writing -- original draft, Writing -- review \& editing. \textbf{Matti Schneider:} Conceptualization, Methodology, Formal analysis, Software, Visualization, Writing -- original draft, Writing -- review \& editing, Supervision, Funding acquisition.

\section*{Declaration of competing interest}

The authors declare that they have no known competing financial interests or personal relationships that could have appeared to influence the work reported in this paper.

\section*{Funding}

FG and MS were funded by the European Union through the European Research Council under the Horizon Europe programme (ERC Starting Grant BeyondRVE, grant agreement No.~101040238). Views and opinions expressed are, however, those of the authors only and do not necessarily reflect those of the European Union or the European Research Council Executive Agency. Neither the European Union nor the granting authority can be held responsible for them.

\section*{Data availability}

The data that support the findings of this study are available from the corresponding author upon reasonable request.

\section*{Declaration of generative AI and AI-assisted technologies in the writing process}
During the preparation of this work, the authors used Claude (Anthropic) to improve the language, to check derivations and numbers, and to assist with coding. No generative AI was used to create or alter images, all figures are generated from data or code. The authors reviewed and edited the content as needed and take full responsibility for the content of the publication.

\appendix
\section{Coefficient matrices of scaled boundary finite element method for 3D elasticity} \label{app:coeff}

As derived in Chapter 6.6 of Song's book~\cite{Song-2018} the matrices ${\mathbf{b}}_i$ in Eq.~\eqref{eq:b1b2b3} are given by the expressions
\begin{subequations}
\begin{equation}
    {\mathbf{b}_1} = \frac{1}{\text{det}({\mathbf{J}})}\left[ \begin{array}{ccc}
      y_{b,\eta} z_{b,\zeta} -z_{b,\eta} y_{b,\zeta}    & 0 &  0\\
      0   & z_{b,\eta} x_{b,\zeta}  - x_{b,\eta} z_{b,\zeta}  & 0 \\
      0 & 0 & x_{b,\eta} y_{b,\zeta}  - y_{b,\eta} x_{b,\zeta}  \\
    0 & x_{b,\eta} y_{b,\zeta}  - y_{b,\eta} x_{b,\zeta}  & z_{b,\eta} x_{b,\zeta}  - x_{b,\eta} z_{b,\zeta}  \\
    x_{b,\eta} y_{b,\zeta}  - y_{b,\eta} x_{b,\zeta}  & 0 & y_{b,\eta} z_{b,\zeta} -z_{b,\eta} y_{b,\zeta}  \\
    z_{b,\eta} x_{b,\zeta} -x_{b,\eta} z_{b,\zeta}  & y_{b,\eta} z_{b,\zeta} -z_{b,\eta} y_{b,\zeta}  & 0
     \end{array} \right],
    \end{equation}
\begin{equation}
    {\mathbf{b}_2} = \frac{1}{\text{det}({\mathbf{J}})}\left[ \begin{array}{ccc}
      z_b \,  y_{b,\zeta} -y_b\,  z_{b,\zeta}   & 0 &  0\\
      0   & x_b\,  z_{b,\zeta}  - z_b x_{b,\zeta} & 0 \\
      0 & 0 & y_b \, x_{b,\zeta}  - x_b\, y_{b,\zeta}  \\
    0 & y_b\,  x_{b,\zeta}  - x_b\,  y_{b,\zeta}  & x_b\,  z_{b,\zeta}  - z_b\,  x_{b,\zeta}  \\
   y_b \, x_{b,\zeta}  - x_b\,  y_{b,\zeta}  & 0 & z_b\,  y_{b,\zeta}  - y_b\,  z_{b,\zeta}  \\
   x_b\,  z_{b,\zeta}  - z_b\,  x_{b,\zeta}  & z_b\,  y_{b,\zeta}  - y_b\,  z_{b,\zeta}  & 0
     \end{array} \right], 
 \end{equation}
 \begin{equation}
  {\mathbf{b}_3} = \frac{1}{\text{det}({\mathbf{J}})}\left[ \begin{array}{ccc}
      y_b \, z_{b,\eta} -z_b\,  y_{b,\eta}   & 0 &  0\\
      0   & z_b\,  x_{b,\eta}  - x_b\,  z_{b,\eta} & 0 \\
      0 & 0 & x_b\,  y_{b,\eta}  - y_b \, x_{b,\eta}  \\
    0 & x_b\,  y_{b,\eta}  - y_b \, x_{b,\eta}  & z_b\,  x_{b,\eta}  - x_b\,  z_{b,\eta}  \\
   x_b\,  y_{b,\eta}  - y_b\,  x_{b,\eta}  & 0 & y_b \, z_{b,\eta}  - z_b\,  y_{b,\eta}  \\
   z_b \, x_{b,\eta}  - x_b \, z_{b,\eta}  & y_b \, z_{b,\eta}  - z_b\,  y_{b,\eta}  & 0
     \end{array} \right],
    \end{equation}
\end{subequations}
where all partial derivatives are marked by a comma in the subscript, following the variable of derivation. We use Mandel notation, i.e., the last three rows of ${\mathbf{b}_1}$, ${\mathbf{b}_2}$, ${\mathbf{b}_3}$ are multiplied by $1/\sqrt{2}$. The Jacobian at the boundary ($\xi=1$) takes the form~\cite[Eq. 6.53]{Song-2018}
\begin{align}
    \text{det}({\mathbf{J}}) = (x_b-x_O) \left(y_{b,\eta} \, z_{b,\zeta} - z_{b,\eta} \, y_{b,\zeta}\right) +(y_b-y_O)\left(z_{b,\eta} \, x_{b,\zeta} - x_{b,\eta} \, z_{b,\zeta}\right)+ (z_b-z_O)\left(x_{b,\eta} \, y_{b,\zeta} - y_{b,\eta} \, x_{b,\zeta}\right)
\end{align}
using the coordinates of the scaling centre $(x_O,y_O,z_O)$ and the coordinates that describe the boundary $(x_b,y_b,z_b)$ of the hexahedral element.

\bibliographystyle{ieeetr}
{\footnotesize \let\oldbibitem\bibitem\renewcommand{\bibitem}{\setlength{\itemsep}{0pt}\oldbibitem}\bibliography{literature}}

\end{document}